\documentclass[fleqn,usenatbib]{mnras}

\usepackage{amsmath}
\usepackage{amssymb}
\usepackage{graphicx}
\usepackage{booktabs}
\usepackage{txfonts}

\newcommand{\rhot}{\rho_t}
\newcommand{\Msun}{\ensuremath{M_\odot}}
\newcommand{\kms}{\ensuremath{\mathrm{km\,s^{-1}}}}
\newcommand{\Vobs}{V_{\mathrm{obs}}}
\newcommand{\Vb}{V_b}
\newcommand{\Vbx}{V_{bX}}
\newcommand{\Vflat}{V_{\mathrm{flat}}}
\newcommand{\rhoo}{\rho_o}
\newcommand{\rhobx}{\rho_{bX}}
\newcommand{\Ups}{\Upsilon}
\newcommand{\at}{a_t}
\newcommand{\azero}{a_0}
\newcommand{\chinu}{\chi^2/\nu}
\newcommand{\gbar}{g_{\mathrm{bar}}}
\newcommand{\gobs}{g_{\mathrm{obs}}}
\newcommand{\Reff}{R_{\mathrm{eff}}}
\newcommand{\Rdisk}{R_{\mathrm{disk}}}
\newcommand{\Rt}{R_t}
\newcommand{\rhoS}{\rho_{\mathrm{S}}}

\title[CCC and the SPARC rotation curves]{Testing Covarying Coupling
Constants (CCC) against the full SPARC rotation-curve sample: a like-for-like
comparison with MOND and NFW}

\author[Author et al.]{%
Rajendra P. Gupta\thanks{E-mail: rgupta4@uottawa.ca} and Nikolaos Samaras
\\
Department of Physics, University of Ottawa, Ottawa, ON, Canada K1N 6N5
}

\date{Accepted XXX. Received YYY; in original form ZZZ}
\pubyear{2026}

\begin{document}
\label{firstpage}
\pagerange{\pageref{firstpage}--\pageref{lastpage}}
\maketitle

\begin{abstract}
The Covarying Coupling Constants (CCC) framework, developed to account for
high-redshift \textit{JWST} observations, contains a mechanism -- a
covarying-constant virtual mass field keyed to local density -- that modifies
galactic dynamics without particle dark matter. We test it against the full
\textit{Spitzer} Photometry and Accurate Rotation Curves (SPARC) sample of 175
disc galaxies, extending an earlier study of a few objects. Working in an
\emph{inverse} formulation, in which each model predicts the baryonic rotation
curve from the observed one, we compare CCC against Modified Newtonian Dynamics
(MOND) and one- and two-parameter Navarro--Frenk--White (NFW) haloes on identical
footing, using the reduced $\chi^2_\nu$. We show that the published
sharp density turn-off in the earlier study is unphysical and replace it with a smooth transition --
the density-space analogue of the MOND interpolating function, introducing no new
parameter. One-parameter smooth-CCC then performs comparably to
galaxy-by-galaxy fitted MOND (the lower $\chi^2_\nu$ in $56$ per cent of galaxies,
with near-equal mean $\chi^2_\nu$ of $2.58$ versus $2.65$; the paired difference is
not statistically significant), while the two-parameter NFW halo shows a
substantially broader fit-quality distribution and a larger tail of poor or
boundary-limited fits (mean $\chi^2_\nu\approx7$).
The CCC turn-off density is not universal (scatter $0.82$ dex) and
correlates with galaxy size, qualitatively consistent with the expected effect of
a spherical reconstruction applied to flattened disc systems. Recast as
an acceleration, however, $a_t = V_{\mathrm{flat}}^2/R_t$ has scatter $0.33$ dex
(on the 91-galaxy resolved subset)
-- matching the MOND scale $a_0$ ($0.34$ dex) -- and comparable magnitude of order
$2\times10^{-10}\,\mathrm{m\,s^{-2}}$, with its size correlation removed. CCC,
though not designed for galactic dynamics, describes rotation curves as well as
galaxy-by-galaxy fitted MOND, while the two-parameter dark-matter halo shows a
broader distribution of fit quality and frequent boundary-limited solutions.
\end{abstract}


\section{Introduction}
\label{sec:intro}

\subsection{Historical development}

Galaxy rotation curves provide one of the most direct dynamical
measurements of the discrepancy between luminous matter and gravitating
mass. Early spectroscopy of M31 already indicated that the enclosed
mass continued to rise beyond the bright central regions
\citep{Babcock1939}. Improved optical and 21-cm observations
subsequently established extended, approximately flat rotation curves
in M31 and other spiral galaxies
\citep{RubinFord1970,RobertsRots1973,RobertsWhitehurst1975,
RubinEtAl1978,RubinEtAl1980}. Radio synthesis observations were
decisive because neutral hydrogen commonly extends well beyond the
bright stellar disc. The resulting curves showed that the orbital
speed often remains nearly constant where a finite luminous disc would
predict a Keplerian decline \citep{Bosma1981,Begeman1989}.

The modern mass-modelling problem may be expressed as
\begin{equation}
\begin{split}
V_{\mathrm{obs}}^{2}(R)
={}&V_{\mathrm{gas}}^{2}(R)
+\Upsilon_{\mathrm d}V_{\mathrm{disc}}^{2}(R)\\
&+\Upsilon_{\mathrm b}V_{\mathrm{bulge}}^{2}(R)
+V_{\mathrm x}^{2}(R),
\end{split}
\label{eq:mass-decomposition}
\end{equation}
where $\Upsilon_{\mathrm d}$ and $\Upsilon_{\mathrm b}$ are the
stellar mass-to-light ratios of the disc and bulge, respectively, $V$ are the velocities, and
$V_{\mathrm x}$ represents either a dark halo or an effective
modification of the gravitational field.

Maximum-disc analyses showed that luminous matter can dominate the
inner regions of high-surface-brightness galaxies, whereas
low-surface-brightness and dwarf systems require a mass discrepancy at
nearly all measured radii
\citep{vanAlbadaEtAl1985,Kent1986,CasertanoVanGorkom1991}. The
luminosity dependence of rotation-curve shape motivated the
universal-rotation-curve programme \citep{PersicEtAl1996}.

High-resolution H$\alpha$, H\,\textsc{i}, and CO observations
subsequently sharpened the cusp--core and rotation-curve diversity
problems. Many dwarf and low-surface-brightness galaxies favour slowly
rising rotation curves and shallow central density distributions, in
contrast to the steep cusps found in dark-matter-only simulations,
although cuspy and intermediate systems also occur
\citep{FloresPrimack1994,Moore1994,deBlokEtAl2001,deBlokBosma2002,
GentileEtAl2004,deBlokEtAl2008,OhEtAl2011,OmanEtAl2015,
SantosSantosEtAl2020}. These observations established that a
successful theory must explain not only the approximate flatness of
outer rotation curves but also their detailed shapes and diversity.

\subsection{The SPARC database}

The Spitzer Photometry and Accurate Rotation Curves (SPARC) database
transformed galaxy mass modelling by combining homogeneous
3.6-$\mu\mathrm{m}$ surface photometry with published H\,\textsc{i}
and H$\alpha$ rotation curves for 175 nearby disc galaxies
\citep{LelliEtAl2016a}. The sample spans morphologies from S0 to
irregular galaxies, approximately five decades in luminosity, and four
decades in surface brightness. The 3.6-$\mu\mathrm{m}$ data reduce,
although they do not eliminate, uncertainty in stellar-mass estimates.
A commonly adopted value is
\begin{equation}
\Upsilon_{3.6}
\simeq
0.5\,
\mathrm{M_{\odot}/L_{\odot}}
\end{equation}
for stellar discs in terms of solar mass $M_{\odot}$ and solar luminosity $L_{\odot}$.

SPARC supplies the observed rotation velocities and the Newtonian
velocity contributions calculated separately for atomic gas, stellar
discs, and bulges. It also provides adopted distances, inclinations,
quality flags, and observational uncertainties. This transparency has
made SPARC a standard benchmark for dark-halo profiles, MOND, modified
gravity, and phenomenological alternatives.

The database nevertheless has limitations. It is neither
volume-complete nor derived from one homogeneous kinematic survey. Its
rotation curves were assembled from observations with different
angular resolutions, sensitivity limits, and reduction methods.
Distance and inclination errors move multiple points from one galaxy
coherently rather than independently. Beam smearing, asymmetric drift,
pressure support, bars, warps, spiral arms, and other non-circular
motions may affect the inferred circular velocity
\citep{TrachternachEtAl2008}. Stellar mass-to-light ratios depend on
stellar populations and the adopted initial mass function, while
molecular, ionized, and optically thick gas may not be fully
represented.

An additional issue is that consecutive velocity measurements are
radially correlated. Recent work has introduced empirical covariance
models for SPARC galaxies and found that allowing radial covariance
broadens inferred halo-parameter uncertainties and can weaken the
apparent preference between competing halo profiles
\citep{ChaseEtAl2026}. Consequently, a point-by-point $\chi^{2}$
analysis with diagonal errors may overstate both goodness-of-fit
differences and parameter precision.

\subsection{Empirical regularities revealed by SPARC}

SPARC strengthened several galaxy-scale relations that any viable
theory must reproduce simultaneously. The baryonic Tully--Fisher
relation (BTFR) connects total baryonic mass $M_{\mathrm{bar}}$ and outer rotation
velocity $V_{\mathrm f}$, approximately as
\begin{equation}
M_{\mathrm{bar}}
\propto
V_{\mathrm f}^{4},
\label{eq:btfr}
\end{equation}
with remarkably small scatter over a broad mass range
\citep{McGaughEtAl2000,LelliEtAl2016b}. Its scatter and weak residual
dependence on galaxy radius provide an important constraint on the
galaxy--halo connection \citep{Desmond2017b}.

The central surface-density relation connects the dynamical central
surface density with that inferred from the stellar distribution
\citep{LelliEtAl2016c}. The universal-rotation-curve approach similarly
finds that appropriately normalized rotation curves exhibit
systematic behaviour with luminosity and morphology
\citep{PersicEtAl1996}. Recent work using extended H\,\textsc{i}
curves in SPARC has investigated whether such universality persists to
approximately twice the optical radius, where the inferred mass
discrepancy is larger \citep{BhatiaEtAl2026}.

The most influential SPARC result is the radial acceleration relation
(RAR), which correlates
\begin{equation}
g_{\mathrm{obs}}(R)
=
\frac{V_{\mathrm{obs}}^{2}(R)}{R}
\label{eq:gobs}
\end{equation}
with
\begin{equation}
g_{\mathrm{bar}}(R)
=
\frac{V_{\mathrm{bar}}^{2}(R)}{R}
\label{eq:gbar}
\end{equation}
for thousands of points from galaxies with widely different masses,
gas fractions, surface brightnesses, and morphologies
\citep{McGaughEtAl2016,LelliEtAl2017a}. At high acceleration,
\begin{equation}
g_{\mathrm{obs}}
\simeq
g_{\mathrm{bar}},
\end{equation}
indicating baryonic dominance. Below a characteristic acceleration of
order
\begin{equation}
g_{\dagger}
\sim
10^{-10}\ \mathrm{m\,s^{-2}},
\end{equation}
the discrepancy increases systematically. Table~\ref{tab:symbols} lists the principal symbols. 

\begin{table}
\centering
\caption{Principal symbols.}
\label{tab:symbols}
\begin{tabular}{ll}
\toprule
Symbol & Meaning \\
\midrule
$\Vobs$, $e_{\Vobs}$ & observed rotation velocity and uncertainty \\
$\Vb$ & baryonic (photometric) velocity, eq.~(\ref{eq:vbar}) \\
$V_{\mathrm{gas,disk,bul}}$ & Newtonian gas/disc/bulge velocity ($\Ups=1$) \\
$\Vbx$ & CCC-predicted baryonic velocity, eq.~(\ref{eq:mbx}) \\
$\Vflat$ & asymptotic (flat) rotation velocity \\
$\Ups_{\mathrm{disk}},\Ups_{\mathrm{bul}}$ & stellar $M/L$ of disc, bulge ($\Msun/L_\odot$) \\
$\rhoo$ & total (``observed'') dynamical density, eq.~(\ref{eq:rhoo}) \\
$\rhobx$ & CCC gravitating baryonic density, $\rho(1-X)^4$ \\
$\rhot$ & CCC turn-off density (free parameter) \\
$X(\mathbf{r})$ & CCC virtual mass field \\
$\Rt$ & turn-off radius, where $\rhoo=\rhot$ \\
$\at$ & CCC turn-off acceleration, $\Vflat^2/\Rt$ (Sec.~\ref{sec:at}) \\
$M_o, M_{bX}$ & enclosed total and gravitating-baryonic mass \\
$\gbar, \gobs$ & baryonic, observed centripetal acceleration $V^2/R$ \\
$\azero$ & MOND acceleration scale \\
$\nu(y)$ & interpolation / smooth-transition function \\
$V_{200}, c$ & NFW virial velocity and concentration \\
$k,\ \nu$ & number of free parameters; degrees of freedom $N-k$ \\
$D, i$ & galaxy distance, inclination\\
\bottomrule
\end{tabular}
\end{table}

The interpretation of the RAR remains contested. Its measured scatter
depends on sample selection, uncertainty propagation, correlated
nuisance parameters, and the adopted stellar mass scale. Analyses have
inferred negligible, small, or measurable intrinsic scatter
\citep{Desmond2017a,StoneCourteau2019}. Bayesian studies also disagree
on whether the data require one universal acceleration scale after
distances, inclinations, and mass-to-light ratios are marginalized
\citep{RodriguesEtAl2018,ChangZhou2019,MarraEtAl2020}. Allowing cold,
unobserved baryonic gas can shift the fitted acceleration scale
\citep{GhariEtAl2019}, while symbolic-regression studies find several
functional forms that describe the SPARC relation comparably well
\citep{StiskalekDesmond2023}. The RAR is therefore a robust empirical
correlation, but its exact functional form, intrinsic width, and status
as a fundamental law remain open questions.

\subsection{Dark-halo interpretations}

In $\Lambda$CDM cosmology, galaxies form inside collisionless
dark-matter haloes. Dark-matter-only simulations predict centrally
cusped, approximately universal profiles, most commonly represented by
the Navarro--Frenk--White form,
\begin{equation}
\rho_{\mathrm{NFW}}(r)
=
\frac{\rho_{\mathrm s}}
{\left(r/r_{\mathrm s}\right)
\left(1+r/r_{\mathrm s}\right)^{2}},
\label{eq:nfw}
\end{equation}
where $\rho_{\mathrm s}$ and $r_{\mathrm s}$ are the characteristic
density and scale radius \citep{NavarroEtAl1996,NavarroEtAl1997}.

Empirical cored profiles, particularly the Burkert form,
\begin{equation}
\rho_{\mathrm B}(r)
=
\frac{\rho_{0}r_{0}^{3}}
{\left(r+r_{0}\right)
\left(r^{2}+r_{0}^{2}\right)},
\label{eq:burkert}
\end{equation}
often provide good descriptions of slowly rising dwarf-galaxy curves
\citep{Burkert1995}. Halo concentration is correlated with halo mass
and assembly history \citep{BullockEtAl2001,DuttonMaccio2014}, while
disc size and rotation depend on angular-momentum acquisition, baryon
retention, and halo response \citep{MoMaoWhite1998}.

The principal issue is not whether flexible halo models can fit
rotation curves; in most cases they can. The stronger question is
whether the resulting halo masses, concentrations, core sizes, and
stellar mass-to-light ratios are consistent with cosmological
predictions. Repeated gas inflows and outflows may cause rapid
potential fluctuations, transferring energy to dark matter and
transforming cusps into cores
\citep{GovernatoEtAl2010,PontzenGovernato2012}. Feedback-dependent
profiles such as DC14 and coreNFW encode this process
semi-analytically \citep{DiCintioEtAl2014,ReadEtAl2016}. Simulations
suggest that core formation is most efficient over a limited range of
stellar-to-halo mass ratio \citep{TolletEtAl2016}.

SPARC fits have tested whether feedback-modified haloes satisfy
expected abundance, concentration, and stellar-mass relations
\citep{KatzEtAl2017}. Hydrodynamical simulations produce RAR-like
correlations through the coupled assembly of baryons and dark matter
\citep{LudlowEtAl2017,KellerWadsley2017,NavarroEtAl2017,
TennetiEtAl2018,DuttonEtAl2019}. Flexible profiles such as
core-Einasto models can also reproduce a broad range of simulated and
observed inner structures \citep{LazarEtAl2020}. Nevertheless, the
detailed intrinsic scatter, the most slowly rising dwarf-galaxy
curves, and the diversity at fixed outer velocity remain sensitive to
feedback prescriptions and observational systematics.

Non-standard dark matter supplies additional mechanisms.
Self-interacting dark matter produces approximately isothermal central
regions whose sizes depend on the self-scattering cross-section, halo
history, and baryonic potential
\citep{KaplinghatEtAl2016,CreaseyEtAl2017,RenEtAl2019}. Recent SPARC
work has incorporated both core-growth and gravothermal core-collapse
solutions with velocity-dependent interactions \citep{JiaEtAl2026}.
Fuzzy or ultralight scalar dark matter generates soliton-like cores,
but galaxy-by-galaxy analyses have difficulty finding one universal
particle mass that fits all SPARC galaxies
\citep{KhelashviliEtAl2023}. These results emphasize that successful
individual fits do not, by themselves, establish a universal
microscopic model.

\subsection{Modified dynamics and gravity}

MOND was constructed to explain galaxy dynamics through a transition
below a universal acceleration $a_{0}$, yielding
\begin{equation}
g_{\mathrm{obs}}
\simeq
\sqrt{a_{0}g_{\mathrm N}}
\label{eq:deep-mond}
\end{equation}
in the deep-MOND regime \citep{Milgrom1983}. The AQUAL
modified-Poisson formulation supplied a non-relativistic field theory
\citep{BekensteinMilgrom1984}. Early rotation-curve studies showed that
extended curves could often be predicted from the baryonic
distribution with the stellar mass-to-light ratio as the principal
galaxy parameter \citep{BegemanEtAl1991,McGaughdeBlok1998}.
Subsequent interpolating functions improved fits across the
Newtonian--MOND transition \citep{FamaeyBinney2005}.

MOND naturally connects asymptotically flat rotation curves, the BTFR,
and the RAR and has produced successful fits to many SPARC galaxies
\citep{LiEtAl2018}. Evidence for an external-field effect in SPARC
galaxies has also been reported, although its significance depends on
environmental estimates and modelling assumptions
\citep{ChaeEtAl2020}. Remaining discriminants include the universality
of $a_{0}$, galaxy clusters, relativistic lensing, large-scale
structure, and the CMB.

Relativistic extensions include TeVeS \citep{Bekenstein2004}. Other
alternatives include scalar--tensor--vector gravity, which has been
fitted to galaxy rotation curves without conventional dark haloes
\citep{MoffatRahvar2013}, and emergent gravity
\citep{Verlinde2017}. The latter does not reproduce all detailed SPARC
rotation-curve shapes without additional freedom
\citep{LelliEtAl2017b}. Bayesian model comparisons remain strongly
prior-dependent. For example, a cored dark-matter profile can be
favoured over a chosen MOND/RAR parametrization for many SPARC
galaxies under specified priors and implementations
\citep{KhelashviliEtAl2024}. Such comparisons test particular
realizations rather than every possible form of dark matter or
modified gravity.

\subsection{CCC and emergent \texorpdfstring{$\alpha$}{alpha}-matter}

The Covarying Coupling Constants (CCC) plus Tired Light (TL) framework (CCC+TL) \citep{Gupta2023} was developed to
address tensions in the standard cosmological model, in particular the abundance and maturity of high-redshift galaxies reported by \textit{JWST}.

The CCC framework was applied to randomly selected seven SPARC galaxies \citep{Gupta2025}. In the cosmological
formulation, correlated evolution of the coupling constants introduces
terms that can be reorganized as effective $\alpha$-matter and
$\alpha$-energy \citep{Gupta2024}. The galaxy application permits the
cosmological parameter $\alpha$ to vary locally with the strongly
inhomogeneous baryonic density and defines
\begin{equation}
X(R)
=
-\frac{\alpha(R)}{H},
\qquad
0\leq X(R)\leq 1.
\label{eq:x-parameter}
\end{equation}

The adopted phenomenology suppresses the effective component in
high-density regions and activates it outside a galaxy-dependent
turn-off density $\rho_{\mathrm t}$. In the published
proof-of-concept, selected SPARC rotation curves are reconstructed
using a spherical approximation. The turn-off density, or equivalently
the turn-off radius, is selected separately for each galaxy so that
the baryonic curve inferred by the CCC prescription approximates the
baryonic contribution tabulated by SPARC \citep{Gupta2025}.

This study is important because it supplies a specific galaxy-scale
proposal within a cosmology containing no particle dark matter. In the
adopted construction, the effective $\alpha$-contributions
vanishes inside the turn-off radius and appears only beyond it. The
proposal therefore avoids placing either a cusped or a cored dark
component at the galaxy centre.

The observed rotation curve is
used to infer the total density; the turn-off density is adjusted
separately for each object; spherical symmetry neglects the disc,
bulge, and gas geometry; and the local function $X(R)$ is derived
to determine baryonic distribution to compare with its observed estimates. It is the inverse of the standard procedure in the literature.

The theoretical programme has a possible route towards derivation of $X(R)$ from first-principle. A covariant scalar--tensor action, the constraint
\begin{equation}
G\propto c^{3},
\end{equation}
scalar-field dynamics, and a screening mechanism have been developed
in the covarying bi-scalar framework
\citep{CuzinattoEtAl2023,CuzinattoEtAl2025}. What remains is to derive
the galaxy-scale solution and its boundary conditions from that
action, rather than determining $X(R)$ through the observed rotation
curve.

\subsection{Requirements for a decisive SPARC test}

A competitive CCC analysis calls for reversing the present reconstruction
pipeline for all models. Starting from $V_{\mathrm{obs}}(R)$ as input, the stellar and gas mass distributions should be calculated, since the former is the kinematically observed quantity whereas the latter is estimated with built-in uncertainty whether all baryon matter has been accounted for. The same universal CCC prescription should then be applied to a quality-controlled SPARC sample and compared, using identical data,
nuisance parameters, and statistical criteria, with NFW and MOND models.

Success requires more than small velocity residuals. CCC must
reproduce the BTFR slope, normalization, and scatter; the RAR and its
residual correlations; and the observed diversity of baryon mass distribution. The
inferred turn-off-density distribution should show a physically
motivated dependence on quantities such as morphology, surface
density, or baryonic composition. If every galaxy requires an
unrelated value of $\rho_{\mathrm t}$, the model would remain
descriptive rather than predictive. Conversely, a narrow or
theoretically derived relation for $\rho_{\mathrm t}$ would constitute
a significant result.

The current literature therefore supports a measured conclusion.
$\Lambda$CDM supplies the most developed framework but requires complex galaxy-formation physics to predict individual rotation curves. MOND provides the most
economical mapping from baryonic distributions to galaxy accelerations
but does not yet have a universally accepted relativistic and
cosmological completion. CCC introduces a potentially unifying
effective component and avoids a central dark-matter cusp in its
present construction, but it has so far demonstrated a limited proof
of concept on a few galaxies. SPARC
provides the appropriate benchmark for moving the CCC proposal from
reconstruction to prediction. This is the primary objective of this paper.

The purpose of the present paper is threefold. First, to extend the CCC analysis
from a handful of galaxies to the full SPARC sample of 175
\citep{LelliEtAl2016a}, with an objective, automated fitting procedure in place of
visual curve-matching. Second, to place CCC on a strictly like-for-like footing
with MOND and with NFW, using the same data, the same fitting direction, the same
error model and the same model-selection criteria, so that any performance
difference reflects the physics rather than the analysis. Third, to identify and
correct a specific simplification in the published CCC prescription -- the sharp
density turn-off -- and to quantify its effect.

We describe the data in Section~\ref{sec:data} and the CCC method, including our
objective fitting and the smooth-transition generalisation, in
Section~\ref{sec:method}. Section~\ref{sec:results} presents the full-sample
results and the model comparison. Section~\ref{sec:physical} examines the
turn-off scale as a physical quantity and its correlations.
Section~\ref{sec:discussion} discusses systematics, limitations and the
interpretation of the parameter scatter. Section~\ref{sec:conclusions} concludes.

\section{Data}
\label{sec:data}

We use the SPARC database \citep{LelliEtAl2016a}, comprising 175 late- and early-type
disc galaxies with \textit{Spitzer} [3.6]-\micron\ surface photometry and
homogeneously-analysed H\,\textsc{i}/H$\alpha$ rotation curves. Of the 175 SPARC galaxies, we retain the 165 with at least six velocity
measurements and a determined flat rotation speed $V_{\rm flat}$; the ten excluded
galaxies have only four or five data points and no measured $V_{\rm flat}$, too
few to reconstruct a density profile or constrain a one-parameter fit. These
$165$ galaxies form the principal rotation-curve comparison; two later
acceleration-based analyses use nested subsets of this sample, namely the $132$
galaxies with a well-defined $\Vflat$ and the $91$ of those with a resolved CCC
turn-off radius $\Rt>R_{\min}$ (Section~\ref{sec:at}). From the
mass-model table (their table~2) we take, at each measured radius $R$: the
observed rotation velocity $\Vobs$ and its uncertainty $e_{\Vobs}$; and the
Newtonian velocity contributions of the gas ($V_{\mathrm{gas}}$), stellar disc
($V_{\mathrm{disk}}$) and bulge ($V_{\mathrm{bul}}$), the latter two computed for
a stellar mass-to-light ratio of unity. From their table~1 we take global
properties: adopted distance, inclination, [3.6] luminosity, effective radius
$\Reff$, disc scale length $\Rdisk$, H\,\textsc{i} mass, the quality flag $Q$, and
the numerical Hubble type $T$ (running from early-type spirals at $T=0$ to
irregulars and blue compact dwarfs at $T=11$; we refer to $T<8$ as spirals and
$T\ge8$ as dwarfs/irregulars).

The baryonic rotation curve is assembled as
\begin{equation}
\Vb^2(R) = V_{\mathrm{gas}}|V_{\mathrm{gas}}|
 + \Ups_{\mathrm{disk}}\,V_{\mathrm{disk}}^2
 + \Ups_{\mathrm{bul}}\,V_{\mathrm{bul}}^2,
\label{eq:vbar}
\end{equation}
where the sign-preserving gas term accommodates the small number of tabulated
negative $V_{\mathrm{gas}}$ values (central holes in the H\,\textsc{i}
distribution). We adopt the SPARC-recommended [3.6] mass-to-light ratios
$\Ups_{\mathrm{disk}}=0.5$ and $\Ups_{\mathrm{bul}}=0.7\,\Msun/L_\odot$
\citep{Schombert2019} as our fiducial values, and treat their uncertainty
explicitly (Sections~\ref{sec:ml} and~\ref{sec:disc-ml}).

Unless otherwise stated, statistical tests use the 165 galaxies with at least six
independent radial points, so that two-parameter fits retain positive degrees of
freedom.

\subsection{Notation}
\label{sec:notation}

\section{Method}
\label{sec:method}

\subsection{The CCC galactic mechanism}
\label{sec:mechanism}

In the spherical approximation the mass enclosed within radius $R$ follows from
the circular velocity as $G M_o(R) = V_o^2(R)\,R$, so the total (``observed'')
mass density is
\begin{equation}
4\pi\rhoo(R) = \frac{1}{R^2}\frac{dM_o}{dR}
= \frac{1}{G}\!\left(\frac{V_o^2}{R^2} + \frac{2 V_o V_o'}{R}\right).
\label{eq:rhoo}
\end{equation}
The CCC cosmology creates the effective mass field $X(R)$ such that the
gravitating baryonic density is $\rhobx = \rho\,(1-X)^4$ while the density
inferred dynamically is $\rhoo = \rho\,(1-X)^2$ \citep{Gupta2025}, where $\rho$ is
the underlying (bare) density. The transition is governed by the local density:
below the turn-off the field is inactive ($X=0$), so $\rhoo=\rhobx=\rho$ and the
dynamics are purely Newtonian; the field activates where the underlying density
falls through the turn-off value, i.e.\ outside the turn-off radius $\Rt$ (defined
by $\rhoo=\rhot$), where the reconstructed density has dropped into the transition
regime. In the active branch the two relations
$\rhobx = \rho\,(1-X)^4$ and $\rhoo = \rho\,(1-X)^2$ give
$(1-X)^2 = \rhoo/\rho = \rhobx/\rhoo$, so that $\rhoo^2 = \rho\,\rhobx$; writing the
bare density at activation as the constant turn-off density $\rho\to\rhot$ then
eliminates $\rho$ and yields the central algebraic relation
\begin{equation}
\rhoo^2 = \rhot\,\rhobx
\quad\Rightarrow\quad
\boxed{\;\rhoo = \sqrt{\rhot\,\rhobx}\;}\qquad (\rhobx < \rhot).
\label{eq:geomean}
\end{equation}
The effective CCC contribution therefore appears \emph{outside} the turn-off
radius, in the low-density outskirts where the local density has fallen through
$\rhot$, and vanishes in the dense inner regions where $\rhoo=\rhobx$.
The gravitating baryonic mass and its rotation curve then follow from
\begin{equation}
M_{bX}(R) = \int_0^R (1-X)^2\,dM_o, \qquad \Vbx = \sqrt{G M_{bX}/R}.
\label{eq:mbx}
\end{equation}
Equation~(\ref{eq:geomean}) is, to our knowledge, a new closed form; it is the
density-space analogue of the deep-MOND relation $g = \sqrt{g_N \azero}$, both
being geometric means with a single constant scale. We exploit this parallel
throughout.

\subsection{Inverse formulation and its justification}
\label{sec:inverse}

The tabulated $\Vobs$ is a direct kinematic measurement with a quantified
uncertainty. The baryonic curve $\Vb$, by contrast, is a \emph{construction} from
photometry and an assumed mass-to-light ratio, carrying both a known component
(the $\Ups$ prior) and an unknown component (unmodelled baryons; component- and
galaxy-dependent $\Ups$; distance and inclination propagation). We therefore
adopt an \emph{inverse} formulation: each model predicts $\Vb$ from $\Vobs$, and
the prediction is compared with the SPARC $\Vb$. This places the cleanly-measured
quantity on the independent axis and the uncertain quantity on the dependent
axis, so that a residual admits a baryon-census interpretation that a forward
($\Vb\to\Vobs$) fit would confound with the model. We verified that in the inverse
direction the CCC field $X(R)$ is constructed entirely from $\Vobs$; the parameter
$\rhot$ is calibrated against $\Vb$ exactly as MOND's $\azero$ or NFW's parameters
are. Unless stated otherwise, all fits in this paper are performed in this inverse direction; we drop the qualifier 'inverse' except where contrasting with the forward direction or where confusion might arise.

\subsection{Objective fitting}
\label{sec:fitting}

\begin{figure*}
\centering
\includegraphics[width=\textwidth]{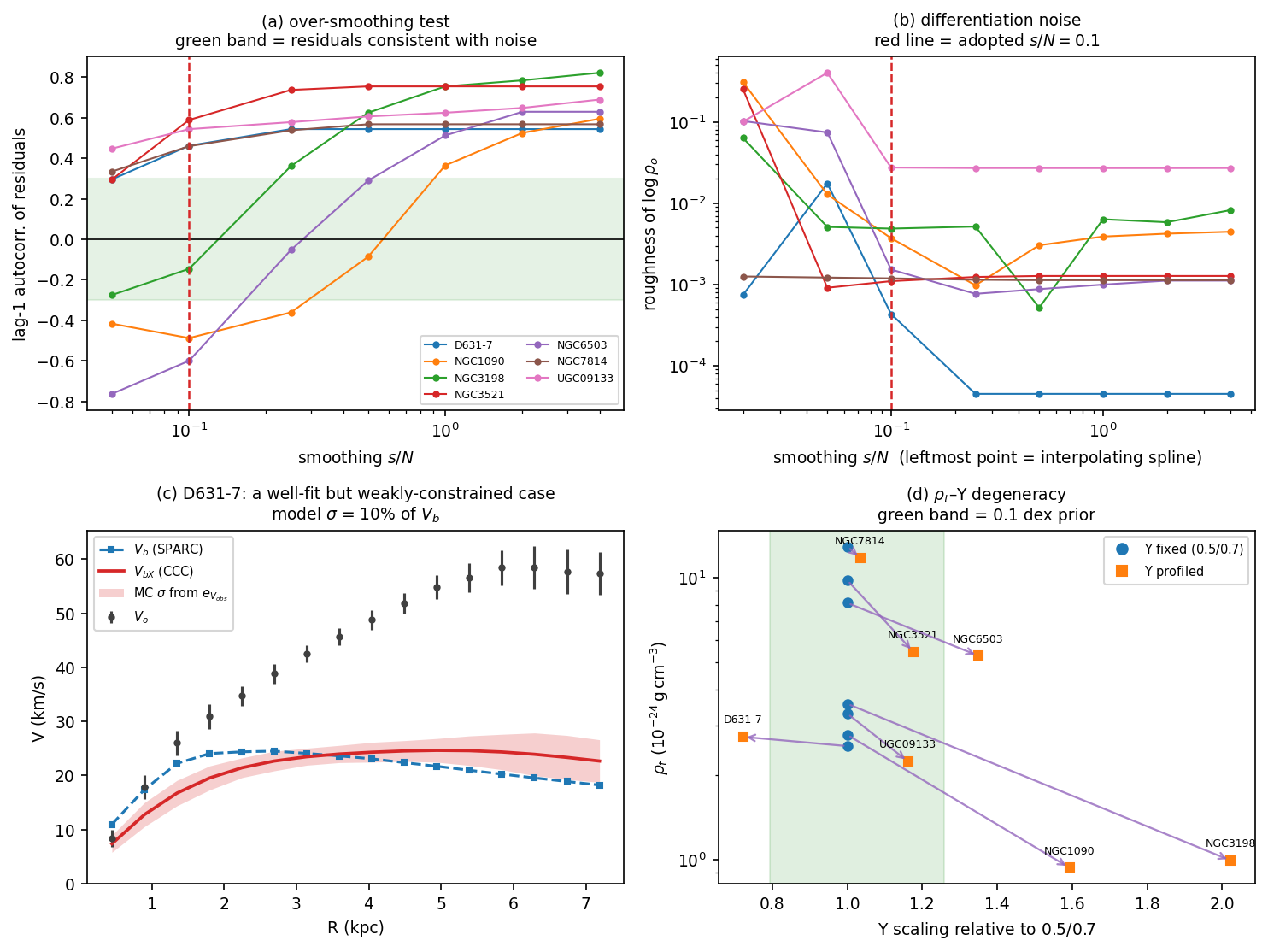}
\caption{Smoothing diagnostics: spline-residual autocorrelation and
reconstructed-density roughness versus smoothing strength, justifying the
adopted $s/N=0.1$.}
\label{fig:diagnostics}
\end{figure*}

The published analysis selected $\rhot$ by visual curve-matching \citep{Gupta2025}. For a sample of
this size we instead minimise 
\begin{equation}
\chi^2 = \sum_i \frac{[\Vbx(R_i) - \Vb(R_i)]^2}{\sigma_i^2}
\label{eq:chi2}
\end{equation}
over the single parameter $\rhot$, by a logarithmic grid scan refined with
Brent's method. The observed density $\rhoo(R)$ is reconstructed by
differentiating an error-weighted smoothing spline fit to $\Vobs(R)$. The
smoothing strength was set by requiring the spline residuals to be consistent
with white noise (lag-1 autocorrelation $\lesssim 0.3$) while suppressing spurious
structure in the reconstructed $d\rho/dR$; a normalised smoothing factor $s/N =
0.1$ satisfies both across the sample (Fig.~\ref{fig:diagnostics}).

We distinguish two senses of ``weighting'' that might otherwise appear to
conflict. Every data point is retained and contributes to every fit: none is
discarded or excluded on the basis of its value or its residual. The spline is
\emph{error-weighted} only in the sense that each point enters with a weight
$1/e_{\Vobs}^2$ set by its \emph{a priori} measurement uncertainty -- the standard
inverse-variance weighting of a least-squares fit -- not by any assessment of how
well it agrees with the model. No point is down-weighted because it is
inconvenient, and no radial range is trimmed; the fiducial results use the
complete rotation curve of every galaxy.

The error model is identical for every model compared. The dominant term is the
propagation of $e_{\Vobs}$ through each model's own inverse chain, evaluated by
Monte Carlo; because the CCC chain (spline $\to$ differentiation $\to$
re-integration) is strongly non-linear, we use the robust 16--84 percentile
half-width of the Monte Carlo ensemble rather than its standard deviation, which
we found to be seed-unstable by a factor of $\sim$3.6. To this we add a 5 per cent
photometric floor on $\Vb$ and a $1\,\kms$ numerical floor.

\subsection{The mass-to-light degeneracy}
\label{sec:ml}

The stellar contribution scales as $\sqrt{\Ups}$, so $\rhot$ and $\Ups$ are
partially degenerate. We treat $\Ups$ as an explicit nuisance parameter with a
$0.1$ dex Gaussian prior \citep{Schombert2019} rather than folding it into the
covariance, having found that a fully-correlated $\Ups$ covariance term silently
absorbs a $\sim$30 per cent normalisation offset and converts the goodness-of-fit
into a shape-only test. Our fiducial results fix $\Ups$ at the SPARC values; a
profiled-$\Ups$ variant is discussed in Section~\ref{sec:disc-ml}.

\subsection{The smooth turn-off (this work)}
\label{sec:smooth}

The published prescription switches sharply from $X=0$ to the active branch at
$\rhoo=\rhot$. This discontinuity is a simplification adopted for analytic
convenience \citep{Gupta2025}; it is unphysical, and its acceleration-space
analogue -- a sharp Newtonian-to-deep-MOND switch -- is known to fit rotation
curves less well than a smooth interpolation. We therefore generalise the CCC
turn-off to a smooth transition acting on the inverse integrand $(1-X)^2 =
\rhobx/\rhoo$. Writing $y=\rhobx/\rhot$ and $\rhoo = \rhobx\,\nu(y)$ with $\nu$ a
monotonic transition function, the integrand becomes $(1-X)^2 = 1/\nu(y)$, which
reproduces the sharp limits ($(1-X)^2\to y$ for $y\ll1$, i.e.\ the active branch
$\rhoo=\sqrt{\rhot\rhobx}$, and $(1-X)^2\to1$ for $y\gg1$, i.e.\ the Newtonian
core) with a smooth knee. We adopt the two standard MOND transition forms,
transcribed into density space.

\noindent\emph{Standard} \citep{McGaughEtAl2016}:
\begin{equation}
\nu_{\mathrm{std}}(y) = \frac{1}{1 - e^{-\sqrt{y}}},
\qquad
\rhoo = \frac{\rhobx}{1 - \exp\!\left(-\sqrt{\rhobx/\rhot}\,\right)}.
\label{eq:std}
\end{equation}

\noindent\emph{Simple} \citep{FamaeyBinney2005,GentileEtAl2004}:
\begin{equation}
\nu_{\mathrm{simple}}(y) = \tfrac12 + \sqrt{\tfrac14 + \tfrac1y},
\quad
\rhoo = \frac{\rhobx}{2} + \sqrt{\Big(\frac{\rhobx}{2}\Big)^{2} + \rhobx\,\rhot}.
\label{eq:simple}
\end{equation}

Both reduce to the sharp published law $\rhoo=\sqrt{\rhot\rhobx}$ in the deep
limit and to $\rhoo=\rhobx$ in the Newtonian limit, and both are continuous at
$\rhobx=\rhot$. We emphasise that this is a \emph{phenomenological} smooth
completion of the sharp CCC turn-off: we transcribe standard monotonic MOND
interpolation forms into density space, and we do not claim that the specific
interpolation function has been derived from the underlying covariant CCC field
equations. The procedure introduces no additional continuous fitting parameter,
retaining $\rhot$ as the single unknown.
The smooth transition is our default throughout; we write simply 'CCC' for smooth-transition CCC and specify 'sharp' for the published work and when required.

\subsection{Comparison models}
\label{sec:models}

\emph{MOND} is represented by the Radial Acceleration Relation (RAR) interpolation
function \citep{McGaughEtAl2016} applied algebraically and pointwise,
$\gobs = \gbar/[1-\exp(-\sqrt{\gbar/\azero})]$, inverted for the inverse-mode
prediction. This matches the pointwise, single-parameter treatment of CCC; we do
not solve the modified Poisson equation. The single parameter is $\azero$.

\emph{NFW} haloes \citep{NavarroEtAl1996,NavarroEtAl1997} have the circular-velocity profile
\begin{equation}
V_{\mathrm{NFW}}^2(R) = V_{200}^2\,\frac{\mu(c\,x)}{x\,\mu(c)},
\quad x \equiv \frac{R}{R_{200}},
\quad \mu(s) \equiv \ln(1+s) - \frac{s}{1+s},
\label{eq:nfw}
\end{equation}
where $R_{200}=V_{200}/(10\,H_0)$ is the virial radius, $c$ the concentration, and
$H_0$ the Hubble constant (we adopt $H_0 = 67.1\,\kms\,\mathrm{Mpc^{-1}}$). The
inverse-mode prediction is $\Vb = \sqrt{\Vobs^2 - V_{\mathrm{NFW}}^2}$. We consider
a one-parameter version (free $V_{200}$, with $c$ fixed by the $c$--$M_{200}$
relation of \citealt{DuttonMaccio2014}) and a two-parameter version (independent
$V_{200}$ and $c$).

All models are fit inverse, with identical $\sigma_i$, identical degrees-of-freedom
accounting, and one or two free parameters as stated.

\subsection{Model selection and statistics}
\label{sec:stats}

We assess fit quality per galaxy with the reduced chi-square $\chinu = \chi^2/\nu$,
where $\nu = N-k$ for $N$ data points and $k$ free parameters. Because $\chinu$ is
normalised per degree of freedom, it is directly comparable across galaxies of
different sizes, and its degrees-of-freedom penalty accounts for the differing
parameter counts of the models we compare. Because the per-galaxy $\chinu$
distribution is strongly non-Gaussian and skewed by a minority of well-sampled
galaxies, we summarise it by the \textbf{mean} across the sample and by
\textbf{head-to-head win fractions}, and we compare the full cumulative
distributions directly (Section 4.5); we avoid the median (which for the
heavily-skewed $\chinu$ distribution can understate the poorly-fit tail) and the
sum (which is dominated by high-point-count galaxies and is not a per-object
measure of quality).

Statistical correlations use the Spearman rank correlation coefficient $\rhoS$
\citep{Spearman1904}, which measures monotonic association between the \emph{ranks}
of two variables and therefore requires neither normality nor linearity and is
robust to outliers -- properties the Pearson product-moment coefficient
\citep{Pearson1895} does not share. We adopt the rank coefficient because both
fitted parameters strongly reject normality under the Shapiro--Wilk test
\citep[][$p\sim10^{-12}$ to $10^{-15}$]{Shapiro1965} and contain outliers up to
$5.5\sigma$, so a Pearson coefficient and its $p$-value would be unreliable. When
testing many predictors simultaneously we control the false-discovery rate by the
\citet{Benjamini1995} procedure, which caps the expected fraction of false
positives among the flagged correlations at $0.05$. Robust linear slopes
(Section~\ref{sec:size}) use the Theil--Sen estimator \citep{Theil1950,Sen1968},
the median of pairwise slopes, insensitive to outliers in either variable.

\section{Results}
\label{sec:results}

\subsection{Validation against the published sub-sample}
\label{sec:validation}

\begin{figure*}
\centering
\includegraphics[width=\textwidth]{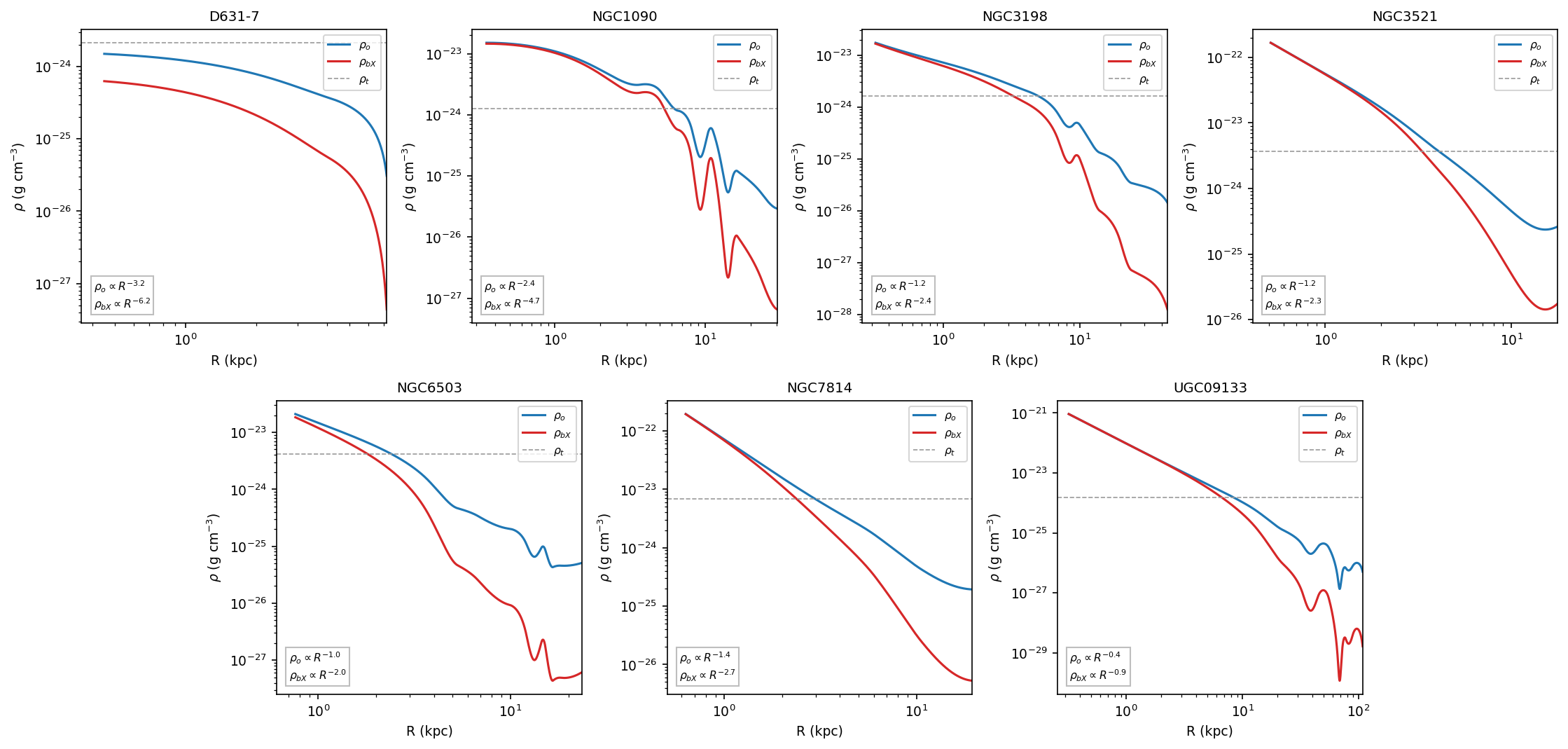}
\caption{Reconstructed density profiles $\rhoo$ and $\rhobx$ for representative
galaxies, showing the predicted $R^{-2}$ and $R^{-4}$ asymptotes.}
\label{fig:density}
\end{figure*}

Applying our objective smooth-transition fit to the seven galaxies of
\citet{Gupta2025} returns turn-off densities that agree with the published values
to within a factor of about two (Table~\ref{tab:validation}). The offset is expected
and systematic: the smooth transition lowers $\rhot$ relative to the published
sharp law by the $\sim$0.6 factor quantified in Section~\ref{sec:sharp}, so our
values lie consistently below the sharp-fit densities rather than scattering about
them. The reconstructed
density profiles reproduce the analytically-predicted asymptotic slopes
$\rhoo\propto R^{-2}$ and $\rhobx\propto R^{-4}$
(Fig.~\ref{fig:density}), a stringent end-to-end check of the differentiation,
inversion and re-integration chain.

\begin{table}
\centering
\caption{The seven \citet{Gupta2025} benchmark galaxies. All quantities in this
work use the smooth transition and the fiducial Monte-Carlo error model, matching
the rest of the paper; the published $\rho_t$ of \citet{Gupta2025} is shown for
reference. $\rho_t$ in $10^{-24}\,\mathrm{g\,cm^{-3}}$, $R_t$ in kpc. A dash marks
an unresolved turn-off ($R_t<R_{\rm min}$).}
\label{tab:validation}
\begin{tabular}{llcccc}
\toprule
Galaxy & Type & $\rho_t$ (2025) & $\rho_t$ (this work) & $R_t$ & $\chi^2_\nu$ \\
\midrule
D631-7   & Im  & 2.79 & 2.13 & --   & 3.67 \\
NGC1090  & Sbc & 2.23 & 1.28 & 6.20 & 1.73 \\
NGC3198  & Sc  & 2.09 & 1.62 & 5.05 & 2.72 \\
NGC3521  & Sbc & 7.96 & 3.67 & 4.10 & 0.08 \\
NGC6503  & Scd & 5.17 & 4.19 & 2.40 & 0.70 \\
NGC7814  & Sab & 5.01 & 6.86 & 2.98 & 0.07 \\
UGC09133 & Sab & 3.98 & 1.56 & 8.32 & 0.32 \\
\bottomrule
\end{tabular}
\end{table}

\subsection{The full sample and the turn-off density distribution}
\label{sec:fullsample}

Figure~\ref{fig:rc} contrasts the smooth and sharp transitions for the seven benchmark galaxies: the sharp law produces a visible kink at $R_t$, which the smooth transition eliminates. Applying the pipeline to all 165 galaxies yields turn-off densities with a median
of $2.9\times10^{-24}\,\mathrm{g\,cm^{-3}}$ and a scatter of $0.82$ dex (standard smooth
transition). The distribution is not consistent with a single universal value; we return
to its physical origin in Section~\ref{sec:physical}.

\subsection{The sharp turn-off is a real but modest deficiency}
\label{sec:sharp}

Fitting MOND with a \emph{sharp} acceleration threshold $g_t$ -- the exact
acceleration-space twin of the sharp CCC law, $\gobs=\sqrt{g_t \gbar}$ for
$\gbar<g_t$ -- quantifies the cost of the sharp transition in a controlled way. In
the inverse direction the sharp threshold is biased $\sim$40 per cent high
relative to the smooth $\azero$ ($g_t/\azero = 1.42$), and $\chinu$ degrades by a
factor $\sim$1.9. The corresponding prediction for CCC -- that a smooth transition
should lower $\rhot$ by a similar factor and improve the fit -- is confirmed below:
the smooth transition returns $\rhot$ at $0.61\times$ the sharp value, explaining a
long-standing normalisation offset between the forward and inverse determinations.

\subsection{Like-for-like model comparison}
\label{sec:comparison}

Table~\ref{tab:ic} gives per-galaxy reduced $\chi^2_\nu$ statistics for all models
in the inverse mode with identical error treatment. We summarise fit quality by
$\chi^2_\nu$, which is normalised per degree of freedom and hence directly
comparable across galaxies of different sizes; because CCC and MOND each carry a
single free parameter and are fitted to the same data, the parameter-count
penalties that distinguish information criteria from $\chi^2$ add essentially the
same constant to each and do not affect the comparison. The smooth CCC transition
brings CCC to parity with MOND: the mean $\chi^2_\nu$ values, $2.58$ (CCC) and
$2.65$ (MOND), are nearly identical, as are their medians ($0.89$ and $1.50$) and
the fractions of galaxies with an acceptable fit ($f(\chi^2_\nu<2)=63$ per cent
for both). The two-parameter NFW halo has both a higher mean ($7.24$) and a much
heavier tail: $24$ per cent of its fits have $\chi^2_\nu>10$, against $7$ per cent
for CCC and MOND, and its $16$--$84$th percentile range ($0.47$--$15.6$) is far
broader than CCC's ($0.21$--$5.1$) or MOND's ($0.50$--$4.1$).

Head-to-head, CCC has the lower $\chi^2_\nu$ in $56$ per cent of galaxies against
MOND; the full cumulative distributions are compared in
Section~\ref{sec:preference} (Fig.~\ref{fig:cdf}). This edge is not statistically
significant: a sign test on the paired per-galaxy differences gives $p=0.16$, a
Wilcoxon signed-rank test $p=0.24$, and a bootstrap $95$ per cent confidence
interval on the mean paired difference, $[-0.54,+0.41]$, includes zero. CCC and
MOND are therefore best described as statistically comparable, not as one
outperforming the other. Against the two-parameter NFW halo, CCC has the lower
$\chi^2_\nu$ in $68$ per cent of galaxies. The NFW difficulty is also visible in
its parameters: of the $165$ two-parameter NFW fits, $25$ ($15$ per cent) reach
the imposed lower concentration boundary $c=1$ and $7$ ($4$ per cent) the upper
$V_{200}=600\,\kms$ boundary (one galaxy reaches both), so $31$ solutions
($19$ per cent) are boundary-limited rather than genuine interior best fits,
the halo being driven to unphysical, near-cored concentrations to accommodate the
data. The fair summary is that CCC and galaxy-by-galaxy fitted MOND have
comparable performance, whereas the two-parameter NFW halo shows a substantially
broader fit-quality distribution and a larger tail of poor or boundary-limited
fits.

We note that the mean $\chi^2_\nu\approx2.6$ for both CCC and MOND sits
somewhat above unity even though the fits are visually good. This is expected and
is not evidence of a poor model: the SPARC formal velocity uncertainties are known
to underestimate the true scatter about a smooth model, since they do not fully
capture non-circular motions, asymmetric-drift and inclination-correction
residuals, or local departures from axisymmetry \citep{LelliEtAl2016a}. Because
these unmodelled contributions are common to every model, they raise the absolute
$\chi^2_\nu$ of all of them roughly equally and cancel in the relative comparison.

\begin{table}
\centering
\caption{Per-galaxy reduced $\chi^2_\nu$ statistics, inverse mode, 165 galaxies
(also where NFW-2p is included), identical error model. Each galaxy's $\chi^2_\nu$
is evaluated separately; lower is better. $k$ is the number of free parameters,
$f_{<2}$ and $f_{>10}$ the fractions of galaxies with $\chi^2_\nu<2$ and $>10$.}
\label{tab:ic}
\begin{tabular}{lccccc}
\toprule
Model & $k$ & mean & median & $f_{<2}$ & $f_{>10}$ \\
\midrule
CCC (smooth, standard $\nu$) & 1 & \textbf{2.58} & 0.89 & 63\% & 7\% \\
CCC (smooth, simple $\nu$)   & 1 & 2.59 & 0.90 & 65\% & 5\% \\
MOND (smooth)                & 1 & 2.65 & 1.50 & 63\% & 7\% \\
CCC (sharp; published)       & 1 & 2.92 & 1.17 & 61\% & 5\% \\
NFW (2-parameter)            & 2 & 7.24 & 2.00 & 50\% & 24\% \\
NFW (1-parameter)            & 1 & 7.81 & 3.48 & 32\% & 24\% \\
\bottomrule
\end{tabular}
\end{table}

\begin{figure*}
\centering
\includegraphics[width=\textwidth]{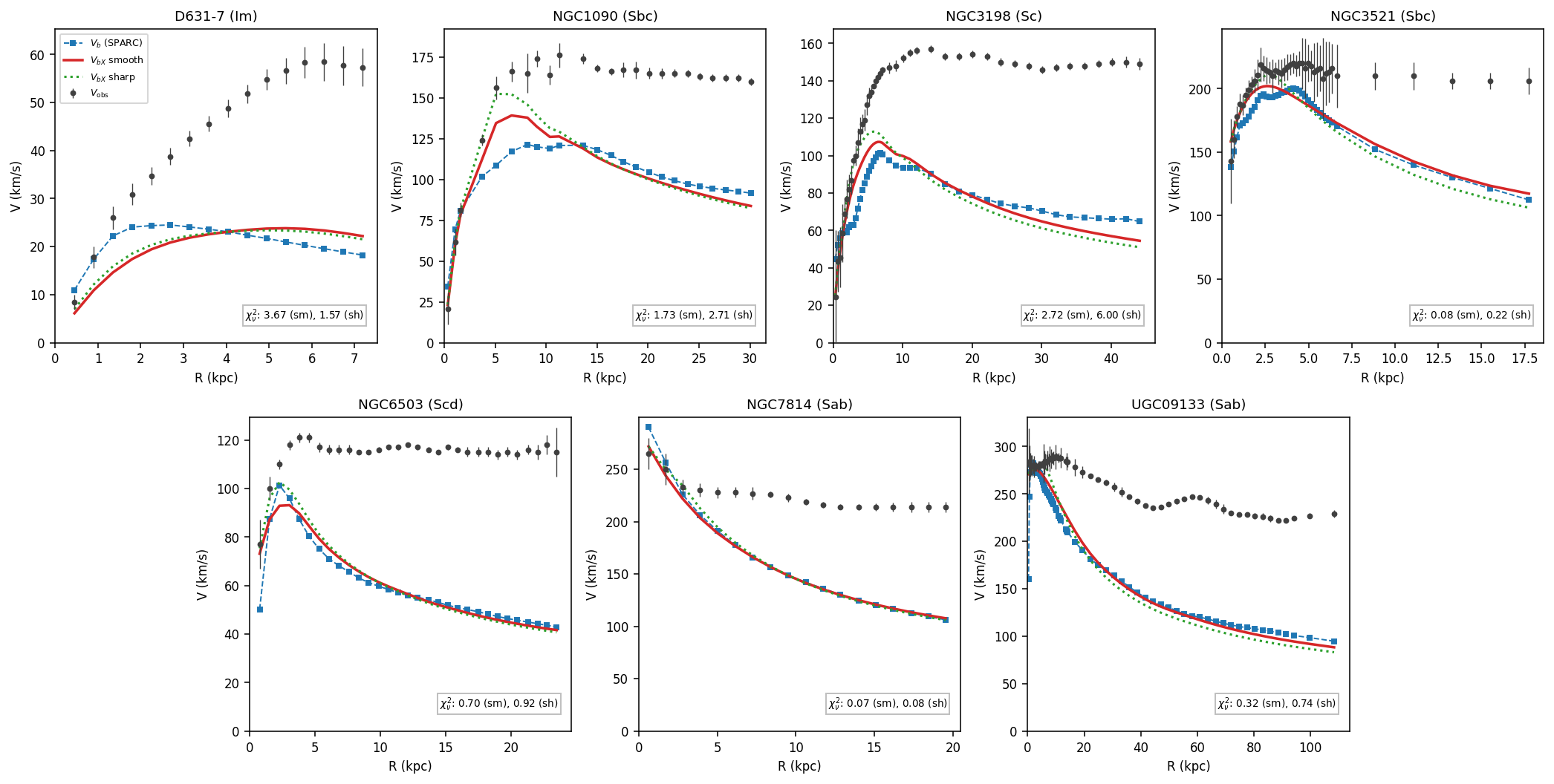}

\caption{CCC fits to the seven \citet{Gupta2025} benchmark galaxies,
spanning early-type spirals (NGC\,7814, Sab) through late-type discs to a gas-rich
irregular (D631-7, Im). In each panel the black points are the observed rotation
velocity $V_{\rm obs}$ with its uncertainty; the blue dashed curve is the SPARC
baryonic velocity $V_b$ (the fit target); the red solid curve is the smooth-CCC
prediction $V_{bX}$ (our fiducial, Eq.~\ref{eq:std}); and the green dotted curve
is the sharp-CCC prediction using the published discontinuous turn-off. The smooth
transition removes the kink the sharp law produces near the turn-off radius,
tracking the baryonic curve more closely (clearest in NGC\,1090 and NGC\,3198).
The per-galaxy reduced $\chi^2_\nu$ is annotated in each panel for both the smooth
(sm) and sharp (sh) transitions. The small $\chi^2_\nu$ of some galaxies is
genuine rather than anomalous: NGC\,7814 ($\chi^2_\nu=0.07$) has a
$V_{bX}$ prediction that hugs the baryonic points across the whole disc, while
NGC\,3521 ($0.08$) has large velocity uncertainties that propagate to a
correspondingly generous prediction error, so agreement within that error yields
$\chi^2_\nu<1$. Both quantities use the fiducial Monte-Carlo error model of
Section~\ref{sec:stats}, identical to that used throughout the paper.}

\label{fig:rc}
\end{figure*}

Figure~\ref{fig:rc} shows the resulting fits for the seven benchmark
galaxies; the full sample is provided as a supplementary montage.

\subsection{The distribution of fit quality, and which galaxies prefer which model}
\label{sec:preference}
\begin{figure}
\centering
\includegraphics[width=\columnwidth]{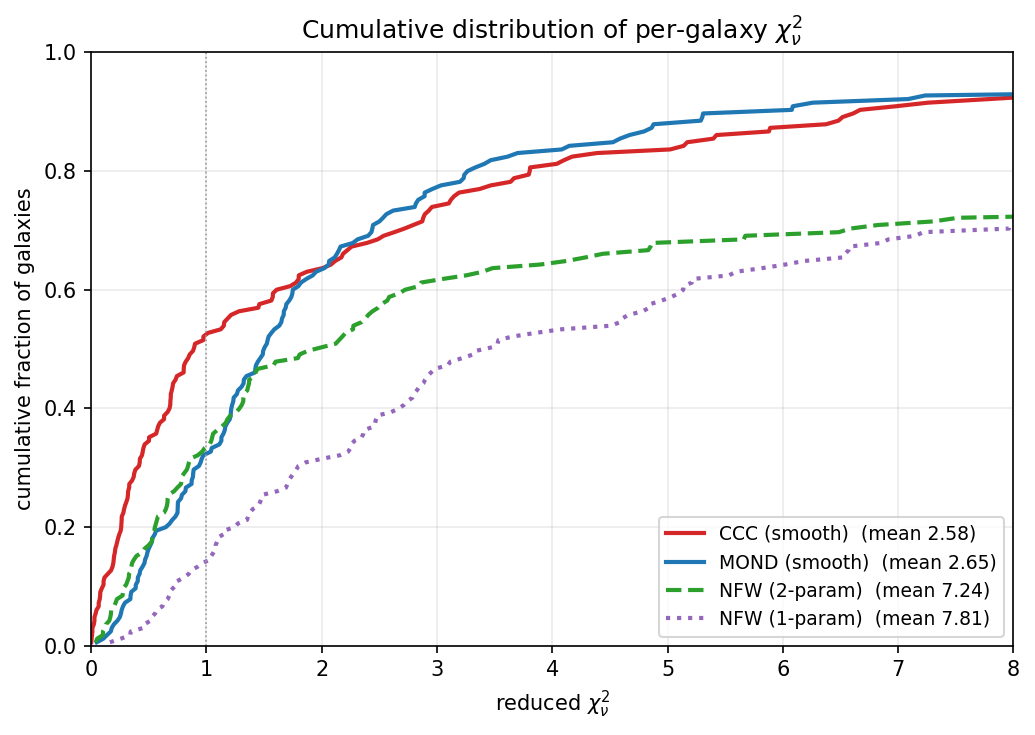}
\caption{Cumulative distribution of the per-galaxy reduced $\chi^2_\nu$ for each
model (after \citealt{LiEtAl2018}). Curves further to the left indicate better fits. CCC and MOND are nearly coincident (their paired difference is not statistically significant), while both show a substantially tighter distribution than the two NFW halo models, whose curves develop long high-$\chi^2_\nu$ tails.
Legend gives the sample-mean $\chi^2_\nu$. We use $\chi^2_\nu$ throughout because it is normalised per degree of freedom and hence comparable across galaxies of different sizes.}
\label{fig:cdf}
\end{figure}

The summary statistics of Section~\ref{sec:comparison} compress each model into a
few numbers; the fuller picture is the \emph{distribution} of per-galaxy fit
quality. Following the approach of \citet{LiEtAl2018}, Fig.~\ref{fig:cdf} shows
the cumulative distribution of the per-galaxy reduced $\chi^2_\nu$ for each model.
This is the cleanest like-for-like comparison available: $\chi^2_\nu$ is
normalised per degree of freedom and so is directly comparable across galaxies of
different sizes, unlike the absolute information criteria, whose per-galaxy values
scale with the number of data points. The CCC curve is better in the lower $\chi^2_\nu$ region, whereas MOND beats CCC where $\chi^2_\nu$ is higher.  While NFW (1 parameter) is no match to either, NFW (2 parameter) is about the same as MOND in the lower $\chi^2_\nu$ region but develops a much heavier tail of poor fits at high $\chi^2_\nu$. Both CCC and MOND are clearly tighter than a dark-matter halo across the bulk of the distribution. The CCC and MOND curves are nearly coincident, consistent with the statistically insignificant paired difference reported above; the two are comparable rather than one being superior.

The per-galaxy $\chi^2_\nu$ values also let us ask \emph{which} galaxies each
model fits better, without invoking the absolute information criteria at all.
Comparing CCC and MOND galaxy-by-galaxy, the sign of the difference correlates
with galaxy properties: CCC tends to fit the high-surface-brightness spirals
better and MOND the gas-rich, low-surface-brightness dwarfs, with a Spearman rank
correlation between the fit-quality difference and effective surface brightness of
$\rho_{\rm S}\approx+0.35$ ($p\sim10^{-5}$; the correlation with flat rotation
speed is very similar). Split by numerical Hubble type, the earlier spirals
($T<8$) favour CCC on average and the later dwarfs and irregulars ($T\ge8$) are
more even or favour MOND.

This division follows the physics of the two prescriptions. CCC keys on the
density, which it reconstructs by differentiating the rotation curve; it therefore
performs best where that curve is steep and well-defined -- the
high-surface-brightness spirals -- and least well where it is shallow and noisy,
as in dwarfs. MOND, mapping acceleration pointwise with no reconstruction, is
robust in exactly the low-surface-brightness regime where CCC's differentiation is
hardest. Part of the pattern is a genuine difference in how the mechanisms engage
the data, and part is a data-quality effect, since high-surface-brightness spirals
are also the best-sampled galaxies; these two axes are entangled in SPARC and we
do not claim to separate them fully. The fair statement is that CCC's advantage is
concentrated where its density reconstruction is well posed, which in the present
data means the massive high-surface-brightness discs, while MOND and CCC are
comparable overall.

\subsection{Residual distributions}
\label{sec:residuals}

\begin{figure*}
\centering
\includegraphics[width=\textwidth]{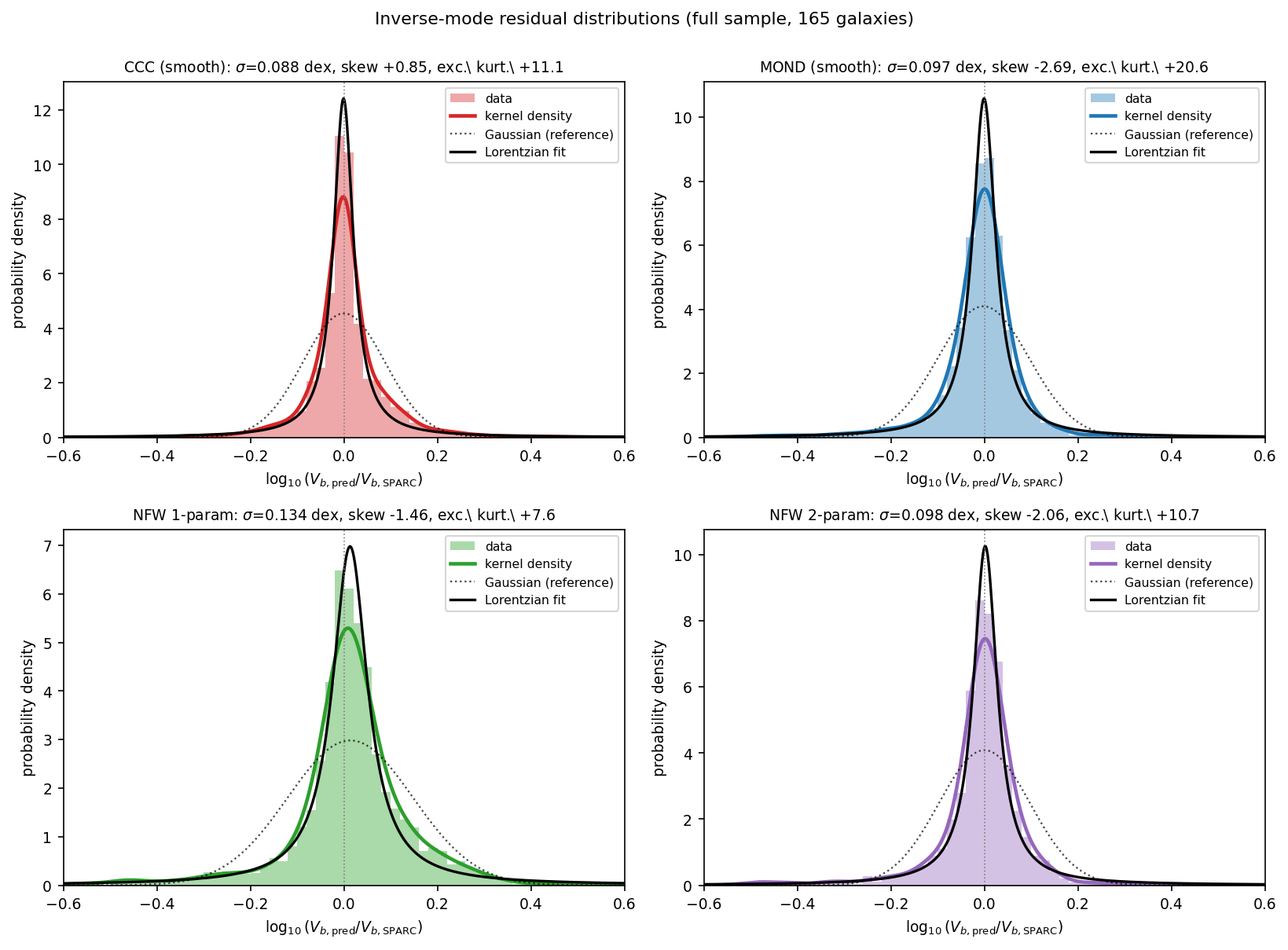}
\caption{Inverse-mode residual distributions
$\log_{10}(V_{b,\mathrm{pred}}/V_{b,\mathrm{SPARC}})$ for all four models. The
filled histogram is the data and the coloured curve its kernel-density estimate;
the dotted curve is a Gaussian of the same median and standard deviation, and the
solid black curve is a Lorentzian (Cauchy) fit. All four distributions are
sharply peaked with heavy tails (excess kurtosis $+7.6$ to $+20.6$): the Gaussian
reference lies below the peak and above the shoulders, whereas the Lorentzian
tracks both the peak and the wings. The Lorentzian is preferred over the Gaussian
for every model by a log-likelihood margin of several hundred
(Table~\ref{tab:resid}). CCC is the only model with a near-symmetric, mildly
positive-skewed residual; the others carry heavy negative tails.}
\label{fig:residuals}
\end{figure*}

The per-point residuals $\log_{10}(V_{b,\mathrm{pred}}/V_{b,\mathrm{SPARC}})$
characterise the models beyond the summary statistics
(Fig.~\ref{fig:residuals}, Table~\ref{tab:resid}). CCC and MOND have
essentially identical widths ($0.088$ vs $0.097$~dex, full sample), with CCC
marginally the tighter and the only model with a near-symmetric, mildly
positive-skewed distribution; MOND and both NFW variants carry heavy negative
tails, reflecting a population of points where they under-predict the baryons in
the high-acceleration inner regions. The one-parameter NFW is the broadest; the
two-parameter NFW matches the CCC and MOND width only by adding a parameter
and at the cost of strongly skewed errors.

All four distributions are sharply peaked with heavy tails, strongly leptokurtic
with excess kurtosis from $+7.6$ (NFW) to $+20.6$ (MOND). Such a shape is poorly
described by a Gaussian, and we find it is instead well fit by a Lorentzian
(Cauchy) profile. For every model the Lorentzian is overwhelmingly preferred over
a Gaussian of the same location and width: the log-likelihood improves by $+556$
to $+1031$, and the Kolmogorov--Smirnov statistic falls by a factor of three to
four (Table~\ref{tab:resid}). The Lorentzian half-widths $\gamma$ are only
$0.026$--$0.046$~dex, quantifying how tightly the bulk of points cluster about a
perfect fit.

This distributional form carries a physical implication, though it does not
uniquely establish its origin. A Gaussian residual distribution is the expectation
when scatter is dominated by many small, independent sources of error --
instrumental noise, thermal broadening, distance and pointing uncertainties --
which the central-limit theorem drives toward a normal distribution. A Lorentzian,
with its power-law wings and formally undefined variance, instead arises when the
scatter is dominated by occasional large excursions rather than many small
independent ones. The observed heavy tails are consistent with a combination of
coherent astrophysical and observational effects, including non-circular motions,
inclination residuals, local departures from axisymmetry, heterogeneous data
quality, and correlated radial errors; the present analysis does not separate
these contributions. Notably this Lorentzian character is shared by all four
models, CCC, MOND and both NFW haloes alike, indicating that it is a property of
the galaxies and the data rather than of any particular gravitational
prescription. It also explains why the summed $\chi^2$ is dominated by a minority
of outlying points (Section~\ref{sec:stats}): the heavy tails are intrinsic to the
data, which is why we summarise fit quality by the mean per-galaxy $\chi^2_\nu$ and
by the cumulative distribution rather than by the sum.

\begin{table}
\centering
\caption{Residual statistics, full distribution. $\sigma$ is the
standard deviation and $\gamma$ the Lorentzian half-width at half-maximum (both in
dex); $\Delta\ln\mathcal{L}$ is the log-likelihood gain of a Lorentzian over a
Gaussian fit (positive favours the Lorentzian), and $D_{\rm KS}$ is the
Kolmogorov--Smirnov statistic against each fitted distribution (smaller is
better).}
\label{tab:resid} 
\begin{tabular}{lcccccc}
\toprule
Model & $k$ & $\sigma$ & skewness & $\gamma$ & $\Delta\ln\mathcal{L}$ &
$D_{\rm KS}^{\rm G}/D_{\rm KS}^{\rm L}$ \\
\midrule
CCC & 1 & 0.088 & $+0.85$ & 0.026 & $+858$  & $0.15/0.04$ \\
MOND & 1 & 0.097 & $-2.69$ & 0.030 & $+1031$ & $0.16/0.05$ \\
NFW (2-param) & 2 & 0.098 & $-2.06$ & 0.031 & $+874$  & $0.17/0.05$ \\
NFW (1-param) & 1 & 0.134 & $-1.46$ & 0.046 & $+556$  & $0.15/0.04$ \\
\bottomrule
\end{tabular}
\end{table}
\section{The turn-off scale as a physical quantity}
\label{sec:physical}

\subsection{A density-space Radial Acceleration Relation}
\label{sec:density-rar}

Unlike the inverse fits used throughout the rest of this work, here we construct
the \emph{forward} density--density relation -- the direct analogue of the
acceleration-space RAR, in which a predicted quantity is built from the baryons
and compared to the observed one. Because equation~(\ref{eq:geomean}) is
pointwise, CCC maps the baryonic density $\rho_b$ to a predicted total density
$\rhoo=\sqrt{\rhot\,\rho_b}$ at each radius, which we compare against the observed
total density inferred from the rotation curve -- a density-space analogue of the
RAR. Constructing this relation symmetrically for both models -- applying the
identical numerical density operator
$\rho=(1/4\pi G)(2g/R+\mathrm{d}g/\mathrm{d}R)$ to each model's predicted curve,
and giving each a single global constant -- yields a scatter of $0.33$~dex for CCC
against $0.23$~dex for MOND.

The gap is partly a differentiation artefact. MOND's native relation lives in
acceleration space, where the pointwise RAR $g_o=\nu(g_b)$ requires no derivative
and has a scatter of only $0.18$~dex. Expressing that same prediction in density
space means passing it through the density operator, whose derivative injects an
additional $0.05$~dex of common-mode scatter, raising MOND to $0.23$~dex. CCC's
relation is intrinsically a statement about densities and cannot be written
without this operator, so it pays the $0.05$-dex ``operator tax'' in full; MOND
pays it only when forced onto CCC's density footing. The remaining difference,
after accounting for the operator, is physical and reflects that CCC keys on the
local density where MOND keys on the enclosed acceleration.

\subsection{The $\rhot$--size relation}
\label{sec:size}

The turn-off density is not universal, but its scatter is partly ordered by galaxy
size. Rank correlations (Table~\ref{tab:corr}) show that $\rhot$ correlates
significantly and \emph{only} with size and mass indicators -- most strongly with
effective radius ($\rhoS = -0.36$, surviving false-discovery-rate control) -- with
a negative sign: larger galaxies have lower $\rhot$. A robust (Theil--Sen) slope
gives $\rhot \propto R^{-\alpha}$ with $\alpha = 0.6$ ($\Reff$) to $0.57$
($\Rdisk$). The correlation survives a partial rank correlation controlling
simultaneously for distance and linear resolution (partial $\rhoS=-0.26$,
$p<10^{-3}$).

Here the partial rank correlation is the Spearman correlation between $\rhot$ and
$\Reff$ after the (rank-transformed) linear dependence of both variables on
distance and resolution has been regressed out; a value of $-0.26$ with $p<10^{-3}$
means the size--density relation retains a monotonic strength of $|\rhoS|=0.26$,
significant at better than the $0.1$ per cent level, once those two potential
confounders are removed. The relation therefore does not arise from the mild
covariance of galaxy size with distance or with linear resolution in the sample.

This is the signature predicted by the projection of a disc density onto a
spherical estimator \citep{Gupta2026}: under
the spherical approximation the estimated density scales as $\Sigma/2R$ against a
true midplane $\Sigma/2h_z$, so $\rhot^{\mathrm{est}}/\rhot^{\mathrm{true}} =
h_z/R$. With SPARC's adopted sublinear disc-thickness relation $h_z\propto
R_d^{0.63}$ this predicts $\alpha\approx0.37$; the measured $\alpha\approx0.6$ is
somewhat steeper, indicating an additional size dependence beyond the thickness
scaling. 

The size-projection effect also accounts for part of the difference between the
full-sample $\rhot$ scatter ($0.82$ dex) and the value on the resolved sub-sample
of Section~\ref{sec:at} ($0.48$ dex): the poorly-resolved galaxies that inflate the
full-sample scatter are preferentially small dwarfs, in which both the projection
effect and the reconstruction noise of Section~\ref{sec:dataquality} are largest.

\begin{table}
\centering
\caption{Spearman rank correlations of the fitted parameters with galaxy
properties (165 galaxies). $\dagger$ survives Benjamini--Hochberg FDR (False Discovery Rate) control at $0.05$.}
\label{tab:corr}
\begin{tabular}{lcc}
\toprule
Property & $\rhoS(\rhot)$ & $\rhoS(\azero)$ \\
\midrule
Effective radius $\Reff$    & $-0.36^\dagger$ & $+0.05$ \\
Disc scale length $\Rdisk$  & $-0.35^\dagger$ & $+0.07$ \\
Gas mass $M_{\mathrm{gas}}$ & $-0.33^\dagger$ & $+0.11$ \\
Baryonic mass $M_{\mathrm{bar}}$ & $-0.31^\dagger$ & $+0.11$ \\
Stellar mass $M_\star$      & $-0.29^\dagger$ & $+0.12$ \\
H\,\textsc{i} radius $R_{\mathrm{HI}}$ & $-0.29^\dagger$ & $+0.13$ \\
Gas fraction $f_{\mathrm{gas}}$ & $+0.19^\dagger$ & $-0.07$ \\
Flat velocity $\Vflat$      & $-0.06$ & $+0.32^\dagger$ \\
Effective surface brightness & $-0.08$ & $+0.16$ \\
Hubble type $T$             & $+0.11$ & $-0.20$ \\
\bottomrule
\end{tabular}
\end{table}

The two parameters correlate with orthogonal quantities: $\rhot$ with size/mass,
$\azero$ with the velocity scale. This supports the interpretation that they are
geometrically inequivalent quantities whose raw scatters are not directly
comparable (Section~\ref{sec:disc-scatter}).

\subsection{The turn-off acceleration: CCC's scale is as universal as $\azero$}
\label{sec:at}

The turn-off can equally be expressed as an \emph{acceleration} rather than a
density, in the same units as MOND's $\azero$. Defining
\begin{equation}
\at \equiv \frac{\Vflat^2}{\Rt},
\label{eq:at}
\end{equation}
with $\Rt$ the radius at which the reconstructed $\rhoo$ crosses $\rhot$, converts
CCC's turn-off scale into a quantity directly comparable with $\azero$. This
requires a resolved turn-off; we therefore restrict the analysis to the 91
galaxies (of 132 with a defined $\Vflat$) for which $\Rt > R_{\min}$, i.e.\ the
turn-off falls within the measured radial range. The remaining galaxies, whose
fitted $\rhot$ lies above the entire reconstructed profile so that $\Rt$ collapses
to the inner boundary, are excluded and discussed as a limitation in
Section~\ref{sec:dataquality}.

On this resolved sample the turn-off acceleration is remarkably universal, and its
scatter matches MOND's (Table~\ref{tab:at}):

\begin{table}
\centering
\caption{Scatter of the CCC turn-off scale in density and acceleration units,
compared with MOND's $\azero$ (resolved sample, 91 galaxies). NMAD is the
normalised median absolute deviation, a robust scatter estimator.}
\label{tab:at}
\begin{tabular}{lccc}
\toprule
Quantity & scatter (dex) & robust (dex) & median \\
\midrule
CCC $\rhot$ (density)      & 0.48 & 0.49 & $2.1\times10^{-24}$\,g\,cm$^{-3}$ \\
\textbf{CCC $\at$ (accel.)} & \textbf{0.33} & \textbf{0.25} & $2.1\times10^{-10}$\,m\,s$^{-2}$ \\
MOND $\azero$ (accel.)     & 0.34 & 0.27 & $9.5\times10^{-11}$\,m\,s$^{-2}$ \\
\bottomrule
\end{tabular}
\end{table}

\begin{figure*}
\centering
\includegraphics[width=\textwidth]{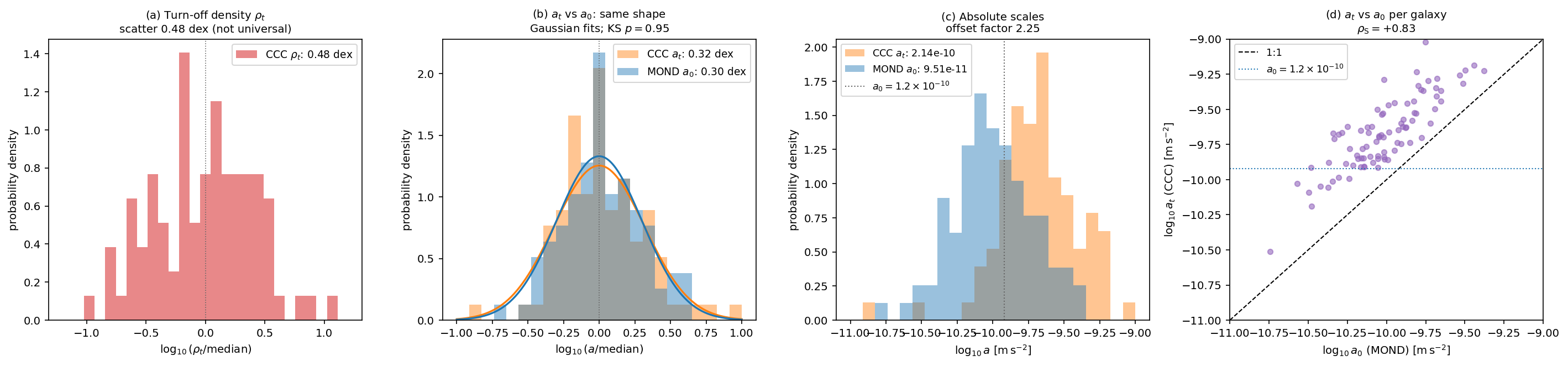}
\caption{The CCC turn-off scale in density and acceleration units (resolved
sample, $N=91$). (a) The turn-off density $\rhot$, normalised to its median, has a
scatter of $0.48$~dex and is not universal. (b) Recast as an acceleration
$\at=V_{\rm flat}^2/R_t$ and centred on their medians, $\at$ and MOND's $\azero$
have statistically indistinguishable distributions (Gaussian fits shown;
two-sample KS $p=0.95$) with nearly equal scatter ($0.32$ vs $0.30$~dex). (c) In absolute terms the two scales differ only by a constant factor of $\sim\!2$; the dotted line marks the canonical $\azero=1.2\times10^{-10}\,\mathrm{m\,s^{-2}}$. (d)
Per galaxy, $\at$ correlates tightly with $\azero$ ($\rho_{\rm S}=0.83$), lying
systematically above the $1{:}1$ line by the geometric offset factor.}
\label{fig:at}
\end{figure*}

Expressed as an acceleration, CCC's turn-off scale is statistically as universal
as MOND's $\azero$ -- $0.33$ against $0.34$ dex, and $0.25$ against $0.27$ dex
robustly -- and of the same order of magnitude, with a median about twice $\azero$
and within a factor of two of the canonical
$1.2\times10^{-10}\,\mathrm{m\,s^{-2}}$ (Fig.~\ref{fig:at}).

The transformation from $\rhot$ to $\at$ also removes the size correlation that
characterises $\rhot$: on this resolved sample, while $\rhot$ correlates strongly
with effective radius ($\rhoS=-0.47$, $p\sim10^{-6}$; the full-sample value in
Table~\ref{tab:corr} is $-0.36$), $\at$ does not ($\rhoS=-0.16$, $p=0.13$, not
significant). This is direct evidence that the excess scatter of $\rhot$ over
$\azero$ is the geometric size-projection lever arm of a density estimator
(Section~\ref{sec:disc-scatter}), which the factor $\Vflat^2$ divides out:
$\Vflat$ carries the same size scaling that contaminates the spherical density
estimate. In its natural acceleration form, therefore, the CCC turn-off is a
near-universal scale of order $\azero$ -- a result made more striking by the
structural parallel between equation~(\ref{eq:geomean}) and the deep-MOND relation
$g=\sqrt{g_N\azero}$, both geometric means with a single constant scale.

\subsection{Morphological dependence and the bulge control}
\label{sec:morph}

\begin{figure}
\centering
\includegraphics[width=\columnwidth]{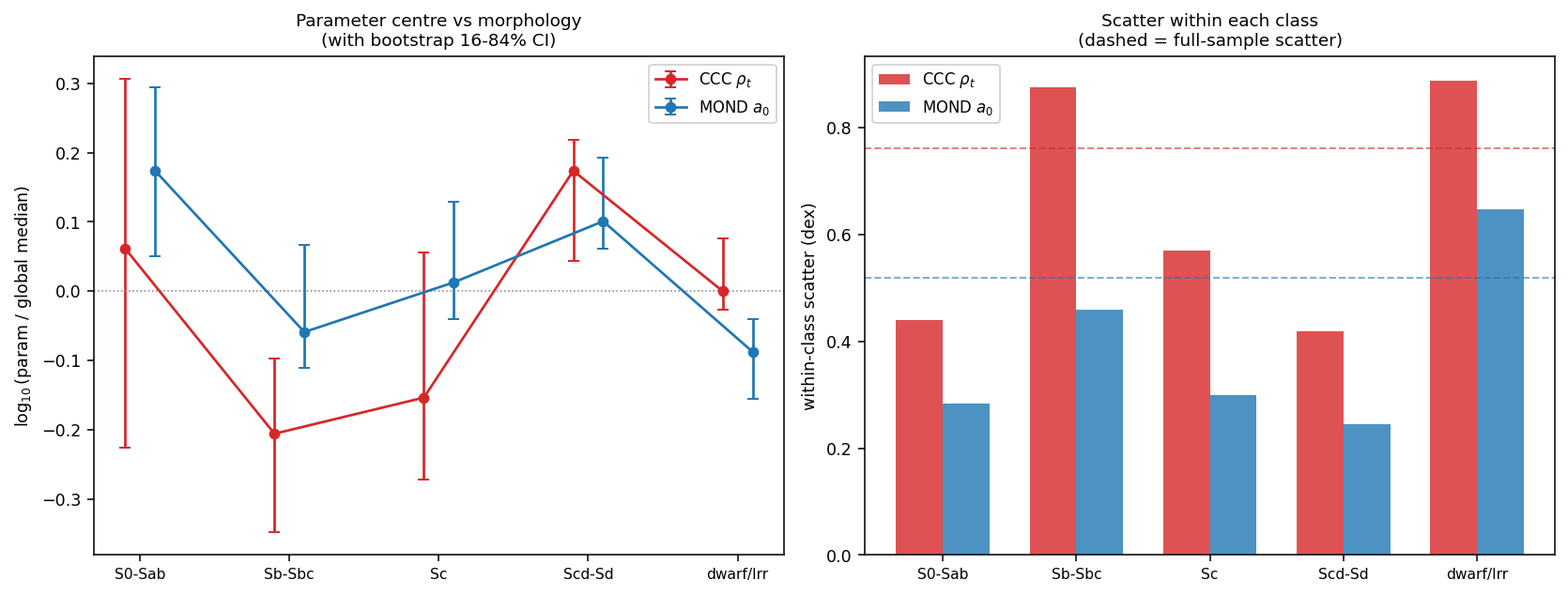}
\caption{Parameter scatter versus morphology. The turn-off density scatter
halves in bulge-dominated galaxies, where the spherical approximation is most
valid; the effect is milder for $a_0$.}
\label{fig:morph}
\end{figure}

The sphericity interpretation predicts that $\rhot$ should be better determined
where the spherical approximation is most valid. This is borne out by the sharpest
available control: bulge-dominated galaxies, whose inner regions are genuinely
near-spherical, have $\rhot$ scatter of $0.42$ dex against $0.82$ dex for
bulgeless galaxies -- a halving. The same split in $\azero$ is milder ($0.29$ vs
$0.55$ dex). A binning by fine Hubble type shows the same qualitative trend but is
not individually resolved bin-by-bin; the bulge/sphericity axis, not the
star-formation morphology, carries the effect (Fig.~\ref{fig:morph}).

\subsection{The baryonic Tully--Fisher relation}
\label{sec:btfr}
CCC makes an analytic prediction for the baryonic Tully--Fisher relation (BTFR),
$M_B\propto V_{\rm flat}^{\,b}$. For $R>R_t$ the geometric-mean law
(equation~\ref{eq:geomean}) gives $(1-X)^2\propto V^2$, and since the gravitating
baryonic mass carries the factor $(1-X)^4$, one obtains $M_B\propto V^4$, i.e.\
$b=4$ \citep{Gupta2025}. We test this empirically, and on equal footing with the
comparison models.

A subtlety of the inverse formulation dictates the correct mass to use. Were we
to plot SPARC's tabulated baryonic mass against the observed $V_{\rm flat}$, the
result would be the model-independent data BTFR, testing nothing about any
model. The model-dependent test instead uses the baryonic mass \emph{each model
infers} from the observed kinematics: in the inverse mode each model predicts a
baryonic velocity curve $V_{b,\rm pred}(R)$, and we take the enclosed baryonic
mass it implies at the outermost measured radius,
$M_B^{\rm model}=V_{b,\rm pred}(R_{\max})^2R_{\max}/G$. We then fit
$\log_{10}M_B = b\log_{10}V_{\rm flat}+\mathrm{const}$ across the sample by the
robust Theil--Sen estimator. This mirrors the published treatment, in which
$V_{\rm flat}$ is a measured quantity and the baryonic mass is model-inferred.

\begin{figure*}
\centering
\includegraphics[width=\textwidth]{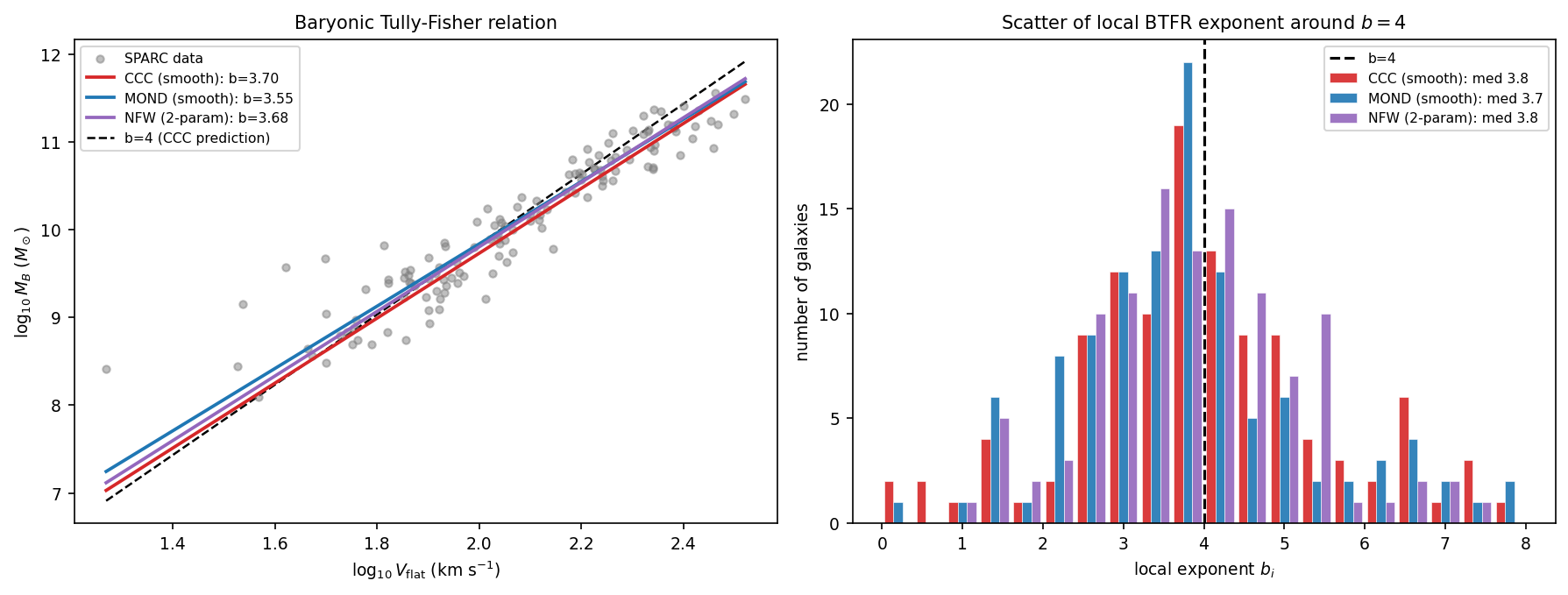}
\caption{Baryonic Tully--Fisher relation for the three models with
model-inferred baryonic mass. Left: $M_B$ versus $V_{\rm flat}$ with Theil--Sen
fits; the dashed line is the CCC-predicted slope $b=4$. Right: distribution of
the per-galaxy local exponent $b_i$; all three models peak near $b=4$.}
\label{fig:btfr}
\end{figure*}

All three models satisfy the BTFR with slopes close to the predicted value and
to one another (Table~\ref{tab:btfr}, Fig.~\ref{fig:btfr}). CCC yields
$b=3.70$, the closest of the three models to the analytic $b=4$; MOND gives
$3.55$ and two-parameter NFW $3.68$, against a data value of $3.45$ from the same
enclosed-mass proxy. The per-galaxy local exponent peaks near $b\simeq3.8$ for
all models. The three models reproduce the data baryonic mass with a rank
correlation of $0.985$ and median offsets below $0.07$ dex, so the BTFR is a
consistency test satisfied by all three rather than a strongly discriminating
one; its value here is in confirming that CCC obeys the relation with the slope
its own analytic argument predicts.

\begin{table}
\centering
\caption{Baryonic Tully--Fisher slope $b$ in $M_B\propto V_{\rm flat}^{\,b}$
(134 galaxies, Theil--Sen; model-inferred $M_B$, inverse mode). The data row uses
the same enclosed-mass proxy and is shown for reference.}
\label{tab:btfr}
\begin{tabular}{lccc}
\toprule
Source & slope $b$ & 95\% CI & scatter (dex)\\
\midrule
SPARC data (proxy) & 3.45 & [3.26, 3.65] & 0.29\\
CCC       & \textbf{3.70} & [3.44, 3.91] & 0.33\\
MOND      & 3.55 & [3.34, 3.76] & 0.30\\
NFW (2-parameter)  & 3.68 & [3.45, 3.89] & 0.34\\
\bottomrule
\end{tabular}
\end{table}

\section{Discussion}
\label{sec:discussion}

\subsection{The inverse formulation}
\label{sec:disc-inverse}

Predicting $\Vb$ from the cleanly-measured $\Vobs$ places the known observational
uncertainty on the independent variable. A residual then admits a baryon-census
interpretation -- a baryon deficiency (unmodelled or under-estimated baryons) or a
baryon excess (over-estimated mass, e.g.\ a mass-to-light ratio set too high or an
over-counted gas mass) -- that a $\Vobs$-target fit would confound with the model.
The sign of the residual distinguishes the two: a deficiency drives the predicted
$\Vb$ above the tabulated value, an excess below it. We stress that this justifies
the \emph{prediction direction} only; we do not claim the residual scatter
\emph{is} a baryon deficiency (or excess). The residual is a signed function of
radius, and its decomposition against local baryon component fractions and radius
shows a mixture: a mild radial trend (projection), a bulge/$M/L$ feature of small
amplitude, and a component common to CCC and MOND in gas-dominated regions that is
therefore attributable to the shared gas inputs rather than to either model. The
gas census itself carries well-known degeneracies (H\,\textsc{i}-to-mass
conversion, optical depth, geometry), so gas-dominated residuals are not a clean
physics probe.

\subsection{The single mass-to-light ratio}
\label{sec:disc-ml}

Our fiducial analysis assumes a universal stellar mass-to-light ratio $\Ups$
(upsilon), with disc and bulge values $\Ups_{\mathrm{disk}}=0.5$ and
$\Ups_{\mathrm{bul}}=0.7\,\Msun/L_\odot$ at [3.6]\,\micron\ (Table~\ref{tab:symbols}).
This is an oversimplification: stellar populations vary within and between
galaxies, and SPARC carries no colour information to set $\Ups$ per galaxy. A
profiled-$\Ups$ variant with a $0.1$ dex prior mildly prefers larger stellar
masses (median scaling $\sim$1.2), but $\Ups$ is weakly constrained by the fit
(16--84 per cent range spanning a factor of two, and $\sim$13 per cent of galaxies
railed against the prior bound), so we report it only as a consistency check, not
a measurement. The assumption should be regarded as an unresolved systematic
affecting all density-based conclusions. As a further check,
Appendix~\ref{app:freeML} reports inverse-mode fits with free ($\Upsilon,D,i$)
under stellar-population priors for a morphology-spanning subset: CCC reproduces
the rotation curves with mass-to-light ratios close to the stellar-population
expectation, whereas MOND matches it on the gas-rich dwarf IC~2574 only by driving
$\Upsilon_{\mathrm{disk}}$ far below any plausible stellar population -- the same
behaviour \citet{LiEtAl2018} report. That appendix also validates our pipeline
against their published fits and confirms that CCC's turn-off acceleration $a_t$
is a genuine characteristic scale, distinct from MOND's $\azero$.

\subsection{Is the parameter scatter comparison meaningful?}
\label{sec:disc-scatter}

MOND's $\azero$ has a smaller raw scatter ($0.52$ dex) than CCC's $\rhot$ ($0.82$
dex). We caution against reading this as a straightforward universality verdict.
Density is a local, differential quantity; acceleration is a cumulative,
integrated one. The CCC parameter therefore has a larger geometric lever arm on
departures from the assumed spherical profile, and part of its excess scatter is
the size-projection effect of Section~\ref{sec:size} rather than intrinsic
non-universality. The orthogonal correlation structure of Table~\ref{tab:corr} --
$\rhot$ tracking size, $\azero$ tracking velocity -- is direct evidence that the
two parameters fail universality in different ways, and that their raw scatters
are not commensurable. The decisive demonstration is
Section~\ref{sec:at}: recast as an acceleration, the CCC turn-off scale
$\at$ has the \emph{same} scatter as $\azero$ ($0.33$ vs $0.34$ dex) and loses its
size correlation. The apparent non-universality of $\rhot$ is thus largely the
geometric consequence of expressing an acceleration-like turn-off in density
units.

\subsection{The data-quality limitation and the two turn-off observables}
\label{sec:dataquality}

CCC's one structural disadvantage relative to MOND is numerical. The inverse CCC
prediction requires differentiating and re-integrating a sparsely-sampled, noisy
rotation curve, an operation MOND's pointwise acceleration mapping does not need.
We confirmed the resulting direction-sensitivity with a control: an NFW halo,
whose inverse is exact algebra, scores identically in both fitting directions,
whereas CCC degrades in the inverse direction where the differentiation is
amplified. This limitation is concentrated in high-surface-brightness spirals with
steeply-declining curves; it caps how tightly CCC can be tested with data of
SPARC's sampling, and is a data-quality rather than a physics limitation.

The same numerical differentiation affects the two turn-off observables -- the
radius $\Rt$ and the density $\rhot$ -- but unequally, and it is worth making the
distinction explicit. The turn-off radius, obtained as a \emph{crossing} of the
reconstructed $\rhoo(R)$ with the level $\rhot$, is locally ill-conditioned: where
$\rhoo$ is steep a small error in the reconstruction displaces the crossing
substantially, and where $\rhoo$ is flat the crossing is ill-defined. For the
$\sim$14 per cent of galaxies whose turn-off is unresolved ($\Rt < R_{\min}$, the
fitted $\rhot$ lying above the entire reconstructed profile) $\Rt$ collapses to the
inner boundary and the derived $\at=\Vflat^2/\Rt$ diverges; these galaxies are
excluded from the $\at$ analysis of Section~\ref{sec:at}. The turn-off
\emph{density}, by contrast, is fixed not by a local crossing but by matching the
\emph{integrated} mass $M_{bX}=\int(1-X)^2\,dM_o$ (equation~\ref{eq:mbx}) to $\Vb$
across all radii, and the integration averages over the point-to-point
differentiation noise -- the property, noted in the original prescription, that
``integration heals the differentiation.'' We verified this directly:
differentiating $\Vb$ to a density and re-integrating recovers $\Vb$ to $0.2$--$0.5$
per cent. Consequently $\rhot$ is buffered against the noise that destroys $\Rt$,
and does not diverge. It is not, however, immune: for the same unresolved galaxies
$\rhot$ is effectively extrapolated below the innermost measured point, where the
finite-difference baseline is shortest and the reconstruction least reliable -- the
regime in which, for example, the innermost density of the dwarf D631-7 varies by
a factor of four with the finite-difference convention alone. This contributes to
the difference between the full-sample $\rhot$ scatter ($0.82$ dex) and the
resolved-sample value ($0.48$ dex; Section~\ref{sec:at}). We regard this as a
data-sampling limitation rather than a deficiency of the model.

For the inner region ($r<R_{\min}$) we adopt as fiducial a power-law
extrapolation that continues the reconstructed density's log--log slope inward,
the slope being measured once from each galaxy's fiducial reconstruction and
held fixed during the Monte-Carlo error propagation so that it propagates
observational noise without acting as a noise amplifier. We also considered a
solid-body inner core ($V\propto r$, constant density); the two prescriptions give
very similar fits (mean $\chinu$ differing by under $0.1$, with neither
systematically preferred) and turn-off densities agreeing to better than $0.05$
dex for $89$ per cent of galaxies, but the solid-body core imposes an unphysical
flat inner density that appears as a
spurious knee at $R_{\min}$ in the reconstructed profiles. Continuing the
observed slope inward is both smoother and marginally better fitting, so we
prefer it. Because this inner reconstruction enters only the CCC prediction and
not the pointwise MOND and NFW fits, the choice does not affect the model
comparison. We do not attempt to determine the inner profile more precisely: any
sharper assumption would replace a transparent data limitation with a
model-dependent extrapolation that would compromise the like-for-like character
of the comparison.

\subsection{Geometry}
\label{sec:disc-geom}

The spherical approximation is applied within the CCC reconstruction. A control in
which the \emph{comparison target} was made spherically consistent left CCC's
residual essentially unchanged, showing that the target-side geometry is not the
origin of CCC's residual; a full three-dimensional reformulation would
additionally change the reconstruction side and is left for future work. We note
that CCC's effective extra mass, following the baryons through
equation~(\ref{eq:geomean}), is disc-like rather than spherical -- a genuine
physical distinction from NFW, shared with MOND, and testable against vertical
dynamics and lensing.

\subsection{Limitations and future work}
\label{sec:limitations}
The present analysis is deliberately restricted to a common inverse formulation
and a spherical CCC reconstruction in order to permit a uniform full-sample
comparison without introducing additional galaxy-specific freedom. Several
extensions are important but are beyond the scope of this work. These include a
full covariance treatment, a forward rotation-curve likelihood with hierarchical
nuisance parameters, a three-dimensional disc--bulge--gas implementation of the
CCC field, and derivation of the smooth local transition directly from the
covariant CCC equations. Mock-galaxy tests of the density-to-acceleration
transformation and independent tests using lensing and vertical dynamics will also
be valuable. These extensions are intended as tests of the physical origin and
predictive power of the CCC galactic mechanism rather than prerequisites for the
present empirical comparison.

\section{Conclusions}
\label{sec:conclusions}

\begin{enumerate}
\item Extended from a handful of galaxies to the full SPARC sample with an
objective fitting procedure, the CCC galactic mechanism describes disc rotation
curves well: with a smooth density turn-off it achieves a mean $\chinu = 2.58$ in
the inverse formulation, comparable to MOND ($2.65$).

\item On a strictly like-for-like footing -- same data, same inverse direction,
same error model, same $\chi^2_\nu$ statistic -- one-parameter CCC is
statistically comparable to galaxy-by-galaxy fitted MOND (the lower $\chi^2_\nu$ in
$56$ per cent of galaxies, mean $\chi^2_\nu$ of $2.58$ versus $2.65$; the paired
difference is not statistically significant). The two-parameter NFW halo shows a
substantially broader fit-quality distribution ($24$ per cent of fits with
$\chi^2_\nu>10$, against $7$ per cent for CCC and MOND) and frequent
boundary-limited solutions ($19$ per cent), with CCC giving the lower $\chi^2_\nu$
in $68$ per cent of galaxies.

\item The published sharp density turn-off is an unphysical simplification.
Replacing it with a smooth transition -- the density-space analogue of MOND's
interpolating function, introducing no new parameter -- is what brings CCC to
parity with MOND, and simultaneously resolves a normalisation offset in the fitted
$\rhot$.

\item The turn-off density is not a universal constant (scatter $0.82$ dex), but
its scatter is partly geometric: $\rhot$ correlates robustly with galaxy size
($\rhoS=-0.36$; $\rhot\propto R^{-0.6}$), qualitatively consistent with the
expected effect of applying a spherical reconstruction to flattened disc systems,
and its scatter halves in bulge-dominated galaxies where the spherical
approximation is most valid. MOND's $\azero$, by contrast, correlates with the
velocity scale, not size.

\item Expressed in its natural form as an \emph{acceleration}, $\at =
\Vflat^2/\Rt$, the CCC turn-off scale is as universal as MOND's $\azero$ (scatter
$0.33$ versus $0.34$ dex on the resolved sample) and of comparable magnitude
(median $2.1\times10^{-10}\,\mathrm{m\,s^{-2}}$, of order $\azero$). The size
correlation present in $\rhot$ vanishes in $\at$, confirming that the apparent
non-universality of the density parameter is largely the geometric consequence of
expressing an acceleration-like turn-off in density units.

\item That a framework constructed to address high-redshift \textit{JWST}
cosmology reproduces galactic rotation curves as well as the phenomenology
purpose-built for them is a non-trivial success of the CCC programme. Within the
common inverse reconstruction adopted here, CCC performs comparably to
galaxy-by-galaxy fitted MOND, while the two-parameter NFW model exhibits a
substantially broader distribution of fit quality and a larger population of poor
or boundary-limited fits.
\end{enumerate}

\section*{Acknowledgements}
RPG is thankful to Stacy McGaugh for the communications that helped shape this work, and to Piyush Singhal for multiple discussions.

\section*{Data Availability}
The SPARC data underlying this article are publicly available at
\url{http://astroweb.case.edu/SPARC/}. The derived fit parameters, per-galaxy
reduced $\chi^2_\nu$ values, and analysis outputs are included as supplementary material accompanying this article.

\bibliographystyle{mnras}
\bibliography{references_merged}

\appendix
\section{Supplementary rotation-curve montage}
\label{app:montage}
The full 165-galaxy rotation-curve montage is provided as supplementary material,
six galaxies per page. In each row the three columns are, from left to right,
\textbf{CCC} (red), \textbf{MOND} (blue), and the two-parameter \textbf{NFW-2p}
(purple). The black points are the observed total rotation curve $\Vobs$ with its
uncertainties, which is the input to the inversion; the grey line is the SPARC
baryonic curve $\Vb$, which is the comparison target; and the coloured line is
each model's inverse reconstruction of the baryonic contribution. The coloured
curves should therefore not be read as forward predictions of the black observed
curve. Each panel is annotated with the model's fitted parameter and its
per-galaxy $\chi^2_\nu$: the turn-off density $\rhot$ for CCC, the acceleration
scale $\azero$ for MOND, and the virial velocity $V_{200}$ and concentration $c$
for NFW-2p. NFW-2p panels whose fit reaches an imposed parameter boundary
($c=1$ or $V_{200}=600\,\kms$) are marked ``boundary'', indicating that the halo
has been driven to an unphysical, near-cored configuration rather than a genuine
interior best fit.

\section{Free mass-to-light fits and the turn-off acceleration scale}
\label{app:freeML}
The main comparison holds the stellar mass-to-light ratios $\Upsilon$, distance
$D$ and inclination $i$ fixed at fiducial values for all models, so that no model
gains per-galaxy freedom the others lack (Section~\ref{sec:ml}). This appendix
reports a complementary test in which these quantities are instead allowed to vary
per galaxy under Gaussian priors (widths $0.11$~dex on $\Upsilon$ about the
population values $0.5$/$0.7$, $10$ per cent on $D$, $5^\circ$ on $i$), following
the nuisance-parameter treatment common in the SPARC literature
\citep{LiEtAl2018}. Two questions motivate it: whether our pipeline reproduces
published fits, and whether CCC can match MOND \emph{without} the unphysical
mass-to-light ratios that MOND sometimes requires.

\emph{Pipeline validation.} Adopting the exact nuisance values of
\citet{LiEtAl2018} for the two galaxies of their Figs.~1 and 2 and fitting only
the acceleration scale in the forward (direct) direction, we recover
$\chi^2_\nu = 1.4$--$1.5$ for both IC~2574 and NGC~2841, close to their quoted
$1.44$ and $1.515$; the fitted $g^\dagger$ returns to $1.2\times10^{-10}\,
\mathrm{m\,s^{-2}}$. Small residual differences are expected, as we fit point
values rather than marginalising over the full posterior. This confirms our
implementation reproduces the established result.

\emph{Free-$\Upsilon$ comparison.} For CCC we replace the fixed $\rhot$ with a
fixed universal turn-off acceleration $a_t = 2.14\times10^{-10}\,\mathrm{m\,
s^{-2}}$ (the median of the resolved-sample $a_t$ distribution,
Section~\ref{sec:at}), converting to each galaxy's $\rhot$ through its
reconstructed profile; this is the CCC analogue of MOND's universal $\azero$. Both
models are then fitted in the inverse direction with free ($\Upsilon,D,i$) under
the priors above. Table~\ref{tab:freeML} gives the results for a
morphology-spanning subset. Across the well-constrained galaxies CCC reaches
comparable or better $\chi^2_\nu$ while requiring mass-to-light ratios close to
the stellar-population expectation; MOND matches CCC on the dwarf IC~2574 only by
driving $\Upsilon_{\rm disk}$ to $0.2$, far below any plausible stellar
population, echoing the $\Upsilon_{\rm disk}=0.07$ that \citet{LiEtAl2018} report
for the same galaxy. We caution that these are point-estimate fits from a
gradient-free optimiser; a Markov-chain Monte Carlo cross-check on NGC~2841
recovers the same parameters and $\chi^2_\nu$ to within the quoted uncertainties,
validating the faster method. We further verified that the two acceleration scales
are not interchangeable: fitting CCC with MOND's $\azero$, or MOND with CCC's
$a_t$, degrades the fits or forces unphysical $\Upsilon$, confirming that $a_t$ is
CCC's own characteristic scale, distinct from and about $1.8\times$ larger than
$\azero$.

\begin{table}
\centering
\caption{Inverse-mode fits with free ($\Upsilon,D,i$) under stellar-population
priors, for a morphology-spanning subset. CCC uses a universal $a_t=2.14\times
10^{-10}\,\mathrm{m\,s^{-2}}$; MOND uses $\azero=1.20\times10^{-10}$. Reduced
$\chi^2_\nu$ is the fit value (prior excluded); parenthesised values are $1\sigma$
uncertainties. Stellar-population synthesis expects $\Upsilon_{\rm disk}\approx
0.5$ at [3.6]\,$\mu$m.}
\label{tab:freeML}
\begin{tabular}{llcccc}
\toprule
Galaxy & Model & $\chi^2_\nu$ & $\Upsilon_{\rm d}$ & $D$/Mpc & $i$/deg \\
\midrule
NGC\,2841 & CCC  & 0.41 & 0.73(1) & 17.8(3) & 81(1) \\
          & MOND & 1.35 & 0.81(3) & 15.9(2) & 79(1) \\
NGC\,7814 & CCC  & 0.22 & 0.73(2) & 22.9(3) & 90(0) \\
          & MOND & 0.98 & 0.89(3) & 18.1(3) & 90(0) \\
NGC\,3198 & CCC  & 1.50 & 1.17(1) &  8.9(1) & 64(1) \\
          & MOND & 1.92 & 0.74(1) & 11.1(2) & 69(1) \\
NGC\,6503 & CCC  & 2.12 & 0.54(1) &  5.8(1) & 72(1) \\
          & MOND & 2.75 & 0.41(1) &  6.6(1) & 77(1) \\
IC\,2574  & CCC  & 1.13 & 0.64(2) &  4.4(1) & 79(3) \\
          & MOND & 0.50 & 0.20(2) &  3.1(1) & 71(2) \\
\bottomrule
\end{tabular}
\end{table}

\bsp
\label{lastpage}
\end{document}